\documentclass{sig-alternate}
\usepackage{hyperref}
\usepackage{subfigure}
\usepackage{graphicx,type1cm,eso-pic,color}

\begin{document}


\title{On the Energy Proportionality of Scale-Out Workloads}

\numberofauthors{1} 
\author{
\alignauthor
Balaji Subramaniam and Wu-chun Feng\\
       \affaddr{Dept. of Computer Science}\\
       \affaddr{Virginia Tech}\\
       \email{\{balaji, feng\}@cs.vt.edu}
}

\maketitle

\begin{abstract}

Our increasing reliance on \emph{the cloud} has led to the 
emergence of \emph{scale-out} workloads. These scale-out workloads are latency-sensitive 
as they are user driven. In order to meet strict latency constraints, 
they require massive computing infrastructure, which consume
significant amount of energy and contribute to operational costs. This
cost is further aggravated by the lack of energy proportionality in
servers. As Internet services become even more ubiquitous, 
scale-out workloads will need increasingly larger cluster
installations. As such, we desire an investigation into the energy 
proportionality and the mechanisms to improve the power consumption of 
scale-out workloads. 

Therefore, in this paper, we study the energy proportionality and power consumption of
clusters in the context of scale-out workloads. Towards this end, we 
evaluate the potential of power and resource provisioning to improve the 
energy proportionality for this class of workloads. Using data serving, 
web searching and data caching as our representative workloads, we first 
analyze the component-level power distribution on a cluster. Second, we 
characterize how these workloads utilize the cluster. Third, we analyze 
the potential of power provisioning techniques (i.e., active low-power, turbo and
idle low-power modes) to improve the energy proportionality of scale-out workloads. 
We then describe the ability of active low-power modes to provide trade-offs 
in power and latency. Finally, we compare and contrast power provisioning 
and resource provisioning techniques. Our study 
reveals various insights which will help improve the energy proportionality and 
power consumption of scale-out workloads. 

\end{abstract}


\section{Introduction}

The proliferation of cloud computing has led to the emergence of 
scale-out workloads~\cite{clearingcloud, cloudsuite}. These scale-out workloads are a part of several 
popular Internet services. For example, Netflix uses data serving mechanisms 
to access vast amount of media~\cite{cassandraweb}, Google uses web search engines 
to index the public Internet in order to respond to search queries~\cite{googlews} 
and Facebook uses data caching mechanisms to access and update
very popular shared content~\cite{fbmc}.

\begin{figure}[h]
\centering
\includegraphics[width=1.0\columnwidth]{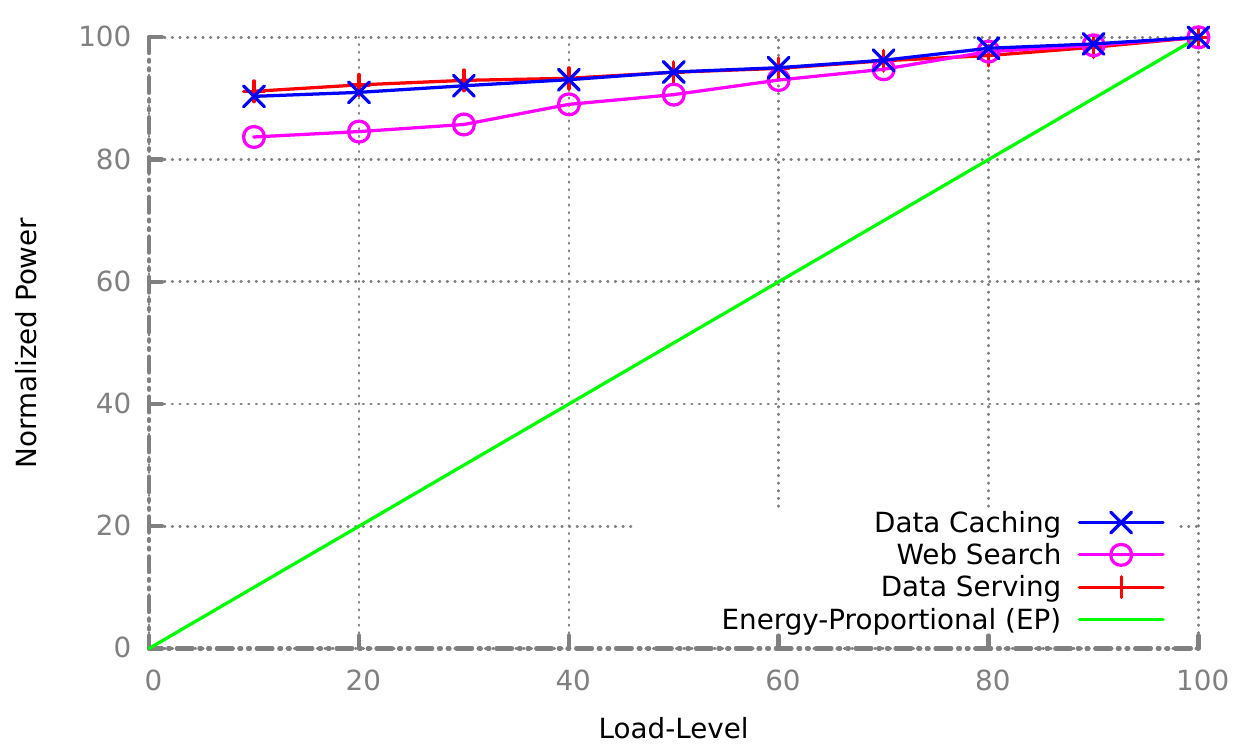}
\caption{Energy Proportionality of Scale-Out Workloads With No Power Management}
\label{fig:epgall}
\end{figure}

Such scale-out workloads are user driven. As a result, they are required to 
meet strict service-level objectives (SLOs), usually in terms 
of sub-second responsiveness. In order to meet the SLO, 
these scale-out workloads can span several hundred to thousands of servers 
to provide efficient access to massive computational resources and huge volumes of data~\cite{wsc}. 
Such installations consume a significant amount of energy, and in turn, contribute to the 
operational costs. Moreover, the operational cost is exacerbated by the lack of energy 
proportionality. Figure~\ref{fig:epgall} shows the normalized 
power consumption of a four-node cluster running scale-out workloads and the 
hypothetical energy-proportional power curve under different load-levels. It 
clearly illustrates the lack of energy proportionality in these workloads.

To address these issues, \emph{power
  provisioning} techniques~\cite{ccgrid_eprop, pegasus, icpe_eprop, coscale, memscale} 
  have been shown to improve the energy proportionality.
These techniques take advantage of low utilization periods or short
idle periods to assign (either active or idle) low-power states to subsystems such as the CPU 
and memory. Other researchers have used \emph{resource provisioning} techniques 
to improve the energy proportionality.
These techniques use workload consolidation to minimize the number of
servers required to sustain a desired throughput and reduce energy
consumption by turning off the servers not in
use~\cite{hpca_knightshift}.

As Internet services such as Netflix, Google and Facebook become even more ubiquitous, 
scale-out workloads will need increasingly larger cluster
installations. Improvements in the energy proportionality of such
installations will need to come from dynamic power management systems. 
Thus, our aim in this paper is to study the energy proportionality of clusters 
in the context of scale-out workloads. Specifically, we analyze the
effectiveness of different software-controlled (hardware-enforced) power management 
techniques, such as \emph{power and resource provisioning}, to improve the energy
proportionality of scale-out workloads. Using a data serving, web search and data caching workload, 
we make the following contributions:

\begin{itemize}
  
\item \emph{A study of the power consumption including power measurements
  of individual components within the cluster.}  We find that
  power consumption is still dominated by the processor. For the scale-out 
  workloads under consideration, our results show that the processor 
  consumes 45-70\% of the system power depending upon the load-level. 

\item \emph{An analysis of the CPU utilization of a cluster executing scale-out workloads.} 
We show that the potential to save power decreases with increase in time resolution and load-level and 
there is ample opportunity to save power even at sub-millisecond granularity. 

\item \emph{An investigation into the energy proportionality improvements, 
the associated performance costs (in terms of response time/latency) and power 
savings achieved using different power provisioning techniques.} Our results show 
that energy-proportional operation is not possible under all load-levels. However, 
improvements in energy proportionality is possible. We also show that 
idle and active low-power together provides the best energy proportionality. We 
save up to 47 and 77 percent of power at the system- and processor-level, respectively. 

\item \emph{An analysis of the power-performance trade-off space exposed 
by the use of active low-power modes for scale-out workloads.} We show that by creating a 
power-performance trade-off space, active low-power modes gives us an opportunity to 
operate a given workload in a power saving configuration while meeting strict SLOs .

\item \emph{A comparison between power and resource provisioning techniques 
including the analysis of the associated performance achieved.} We expose the trade-off 
in power and resource provisioning. We show that 
using resource provisioning at low load-levels provides the best energy 
proportionality as idle power becomes a large portion of the cluster 
power consumption. However, the best power 
saving gradually shifts from resources provisioning (at low load-levels) 
to power provisioning (at high load-levels).

\end{itemize}

The rest of the paper is organized as follows. A brief overview of the 
scale-out workloads, experimental setup, workload configuration, the power measurement interfaces and 
the power management techniques is described in Section~\ref{sec:bkgnd}. We describe the component-level 
power distribution of the scale-out workloads in Section~\ref{sec:powerdis}. Section~\ref{sec:util} 
presents our characterization of the CPU utilization by the scale-out workloads. We show the effects of 
power provisioning techniques on energy proportionality, latency and power savings in Section~\ref{sec:eprop}.
In Section~\ref{sec:pvlat}, we give an overview of the power-performance trade-off space exposed by power provisioning 
techniques. We compare and contrast power and resource provisioning in Section~\ref{sec:ppvrp}. 
A discussion of the related work is presented in Section~\ref{sec:related}. 
Section~\ref{sec:conclusion} concludes the paper.

\section{Background}
\label{sec:bkgnd}

In this section, we present the following background information to
provide context for our work: (1) a description of the scale-out workloads under 
investigation, (2) the experimental setup including the cluster and workload 
configuration, (3) the power management interface and
(4) a brief overview of the power management techniques used in this paper.

\subsection{Scale-Out Workloads}

Here we describe the scale-out workloads under investigation. 

\subsubsection{Data Serving}

NoSQL data stores are popularly used as a data-serving application,
particularly to handle the vast amount of data produced in large-scale web
applications. These data stores provide fast and scalable storage with
unconventional storage schemas. The entire set of data is partitioned
and stored in many different servers.  A key-value store is used to
respond to queries from clients. A middleware layer handles 
the aggregation of the data required for a single client query and the
servers respond to the middleware independently.

\subsubsection{Web Search}
A typical search engine
maintains search indexes that are distributed among several compute nodes (or
index-serving nodes) with each node containing a part of the index
from the Internet. The index-serving nodes are responsible for processing
requests on its own part of the search index. A master node receives
requests from a client, sends requests to all the index-serving nodes,
collects all the responses from them, and sends a appropriate response
back to the client.

\subsubsection{Data Caching}
Data caching seeks to alleviate the load on databases that are used by
large-scale web applications. Performance is a key concern when 
thousands of client requests must be supported simultaneously. A data
cache can improve the performance by decreasing the look-up time
for frequently accessed information. Spare memory in servers is
aggregated to store frequently accessed results of database queries.

\subsection{Experimental Setup}
A four-node cluster is used as the evaluation testbed. 
Each node consists of an Intel Xeon E5-2620 processor, 16~GB of memory 
and a 256-GB hard disk. For all the workloads, a separate server
runs the workload generator or the client. We used the 
workloads from CloudSuite -- a benchmark suite for emerging scale-out 
workloads~\cite{cloudsuite}. Also, we used the latest versions of the software when possible. 
In rest of this section, we describe the software and configurations used 
for each of the scale-out workload under investigation. 

\subsubsection{Data Serving Workload}

We use Cassandra (version 2.0.7)~\cite{cassandra_paper,cassandraweb} as our
distributed NoSQL data store. Cassandra aims to manage large amounts
of data distributed across many commodity servers. It provides a
reliable, high-availability service using a peer-to-peer
architecture. The data is split across each node in the cluster. 

To generate the workload for our experiments, we use the Yahoo! Cloud
Serving Benchmark (YCSB) (version 0.1.4)~\cite{ycsb}. YCSB is a benchmarking framework to evaluate the
performance of cloud data-serving systems. The framework consists of a
load-generating client and a set of standard workloads, such as
read-heavy or write-heavy workloads, which helps in stressing
important performance aspects of a data serving system.

\emph{Configuration:} We load 25 million records into the data store with
a replication factor of three to simulate a realistic setup. The total data stored was 
approximately 80~GB in size (20~GB per server). We evaluate this setup 
using a predefined read-modify-write workload from YCSB. 
The data access pattern follows a Zipfian distribution. The arrival rate of requests follows a 
exponential distribution.\footnote{We use exponential or negative exponential distribution for 
arrival rate of requests for all the scale-out workloads. These are the recommended distributions 
for the arrival rate of requests when real user trace is not available~\cite{oldi}.} The YCSB client reports 
the performance achieved in terms of throughput and latency, 
specifically average latency and latencies at the 95th and 99th percentile. 

\subsubsection{Web Search Workload}

We use Apache Nutch (version 1.2)~\cite{nutchweb} as our web search workload. Nutch is an open source 
web crawler and searcher. It provides a framework for distributed indexing and 
search. For the front-end, Apache Tomcat (version 7.0.23)~\cite{tomcatweb} is used. 

We use Faban toolkit (version 1.0.1)~\cite{faban} as our workload generator. The Faban 
toolkit is a Java-based driver framework which allows for easy definition 
and load generation for new benchmarks. This framework allows the user to 
specify load parameters such as the ramp-up time, steady state time, 
ramp-down time, number of search users and number of client threads. 

\emph{Configuration:} We crawl the public Internet and use an index and segment size of approximately 
1~GB and 14~GB, respectively, per server. The Faban driver uses a negative 
exponential distribution for the arrival rate of requests~\cite{fabandesign}. The Faban 
driver uses latency and throughput as the performance metric. It reports 
90th and 99th percentile latencies. 

\subsubsection{Data Caching Workload}

We use Memcached~\cite{memcachedweb} as our data caching workload. Memcached is a distributed 
in-memory key-value store for generic data. It is used to store popularly 
used data from database queries or page accesses. 

The Memcached client provided with the CloudSuite is used as our workload 
generator. This client allows the user to specify various parameters such 
as ratio of get and set operations. 

\emph{Configuration:} We store approximately 15GB of data per node. A 
scaled version of the Twitter dataset provided with the CloudSuite is used.
We use an exponential arrival rate distribution. The Memcached client 
reports performance in terms of throughput and latency (90th, 95th and 
99th percentile latencies are reported). 

\subsection{Power Measurement Interface}

To study energy proportionality in the context of scale-out workloads, we 
measure the power consumption while executing these workloads at different 
load-levels. We use a \emph{Watts Up} power meter to measure the cluster power 
consumption. Intel's Running Average Power
Limit (RAPL)~\cite{intelsdm,rapl} interfaces is used to measure power of 
components within a system.

\subsection{Power Management}

We study four types of power management techniques in this 
paper. These techniques can be classified into two broad categories: power 
provisioning and resource provisioning. A brief overview of these categories 
is given below. 

\subsubsection{Power Provisioning}
\label{sec:bkgndpprov}
Power provisioning can be further divided into three power management 
techniques. We describe these mechanisms here. 

\emph{Active Low-Power Modes:} These low-power modes are the most popularly used 
form of dynamic voltage frequency scaling (DVFS). It is also referred as P-state 
(or performance state). They are voltage-frequency 
pairs which allow the processor to be operated at lower than the rated 
frequency. When the frequency along with voltage is reduced, the operational 
speed and the power consumption of the processor reduces. In this paper, 
we use Intel RAPL to manipulate active low-power modes. RAPL interfaces 
provides mechanisms to limit the power consumption of the processor.\footnote{Internally 
RAPL uses DVFS.} RAPL interfaces can be programmed using model specific registers 
(MSRs). We refer the reader to the Intel software developer's manual~\cite{intelsdm} 
and existing literature~\cite{barrypowerlimit, icpe_eprop} for more information on 
RAPL interfaces. The rated and minimum frequency of the processors in our experimental setup is 
2GHz and 1.2GHz, respectively. The rated and the minimum frequency are referred as 
$P1$ and $Pn$ states ($n=9$ in our experimental setup)~\cite{snbpower}. In other words, 
there are nine voltage-frequency pairs at which the processor can operate. 

\emph{Turbo Mode:} This is another form of DVFS. This mode overclocks the 
processor above the rated frequency over short duration to improve 
performance. This mode is hardware-controlled and the frequency\footnote{This frequency is above the 
rated frequency.} at which the processor operates depends upon the thermal and power headroom 
available to the processor. In our cluster, the processor can operate at a maximum frequency of 2.5GHz 
when in turbo mode. This maximum turbo frequency is referred as $P0$ state~\cite{snbpower}. 

We disable and enable turbo mode by setting and clearing bit 32 of \emph{IA32\_PERF\_CTL} MSR. 
We also ensure that the appropriate P-states are used by writing the corresponding 
frequencies to the \emph{/sys/devices/system/cpu/cpux/cpufreq/\\scaling\_max\_freq} 
and \emph{/sys/devices/system/cpu/cpux/\\cpufreq/scaling\_min\_freq}. 

\begin{figure*}[t]
\centering
\includegraphics[width=.329\linewidth,height=4cm]{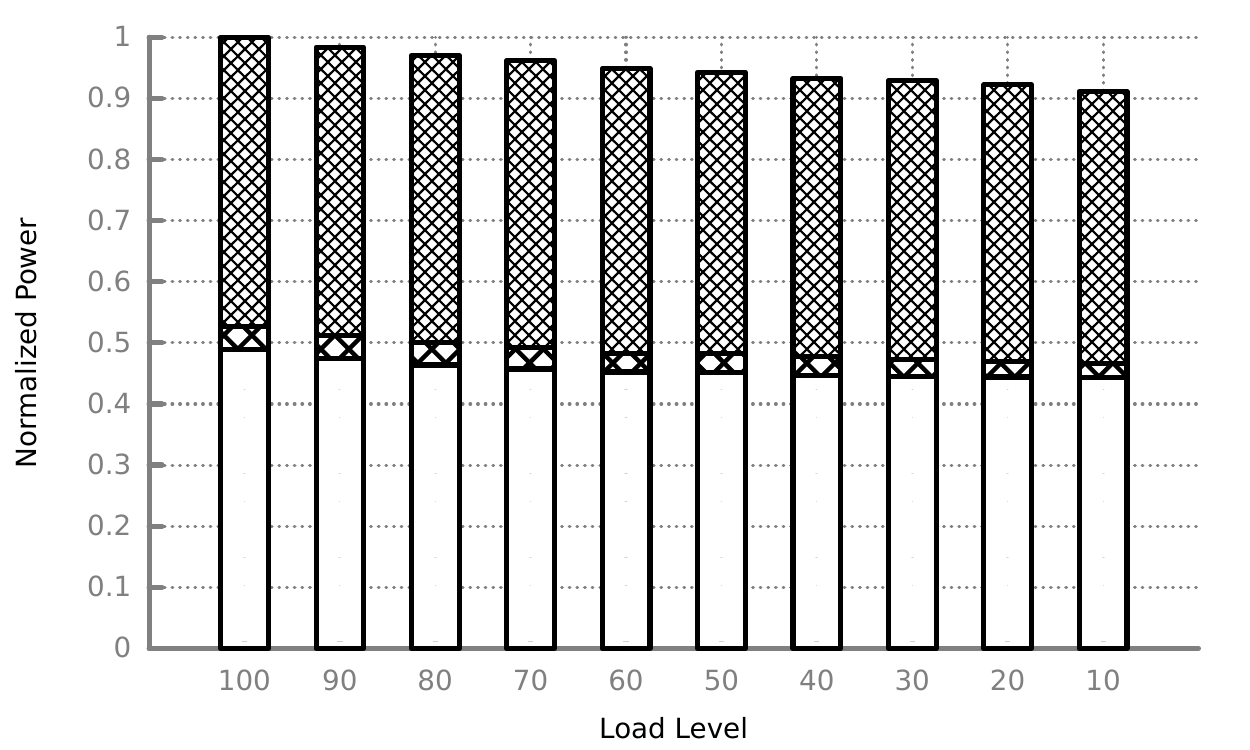}
\includegraphics[width=.329\linewidth,height=4cm]{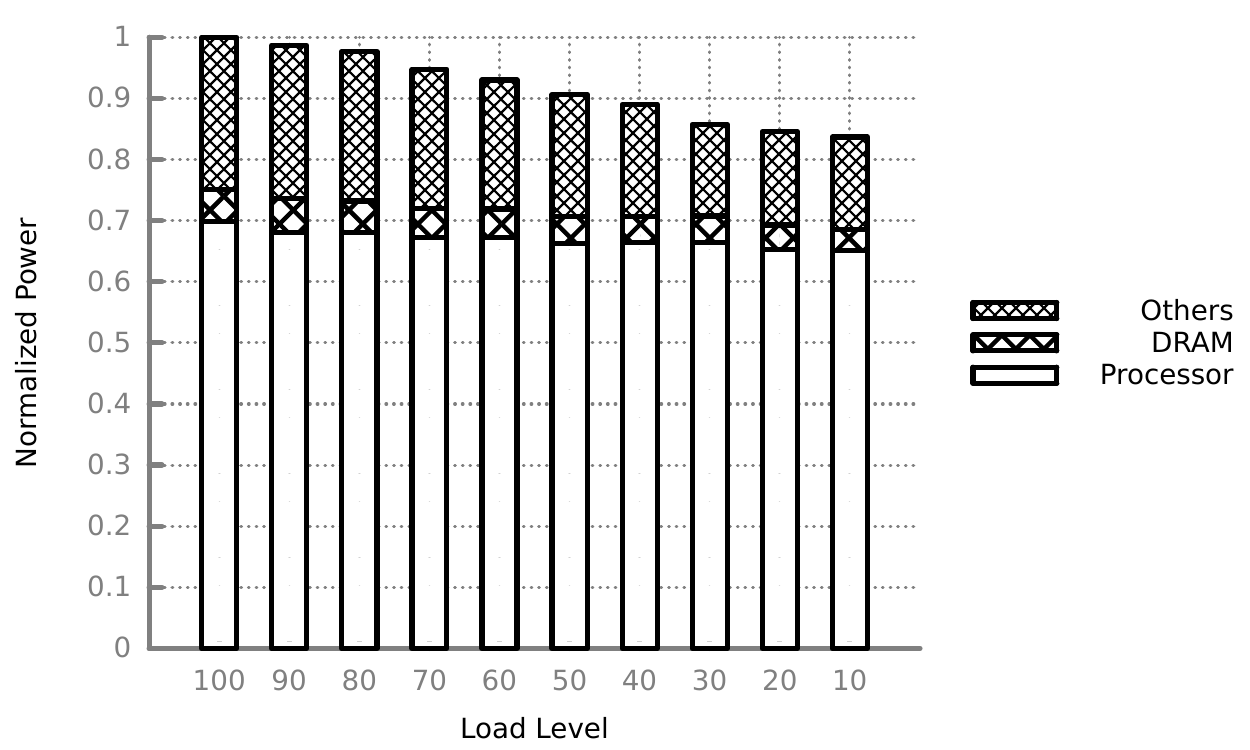}
\includegraphics[width=.329\linewidth,height=4cm]{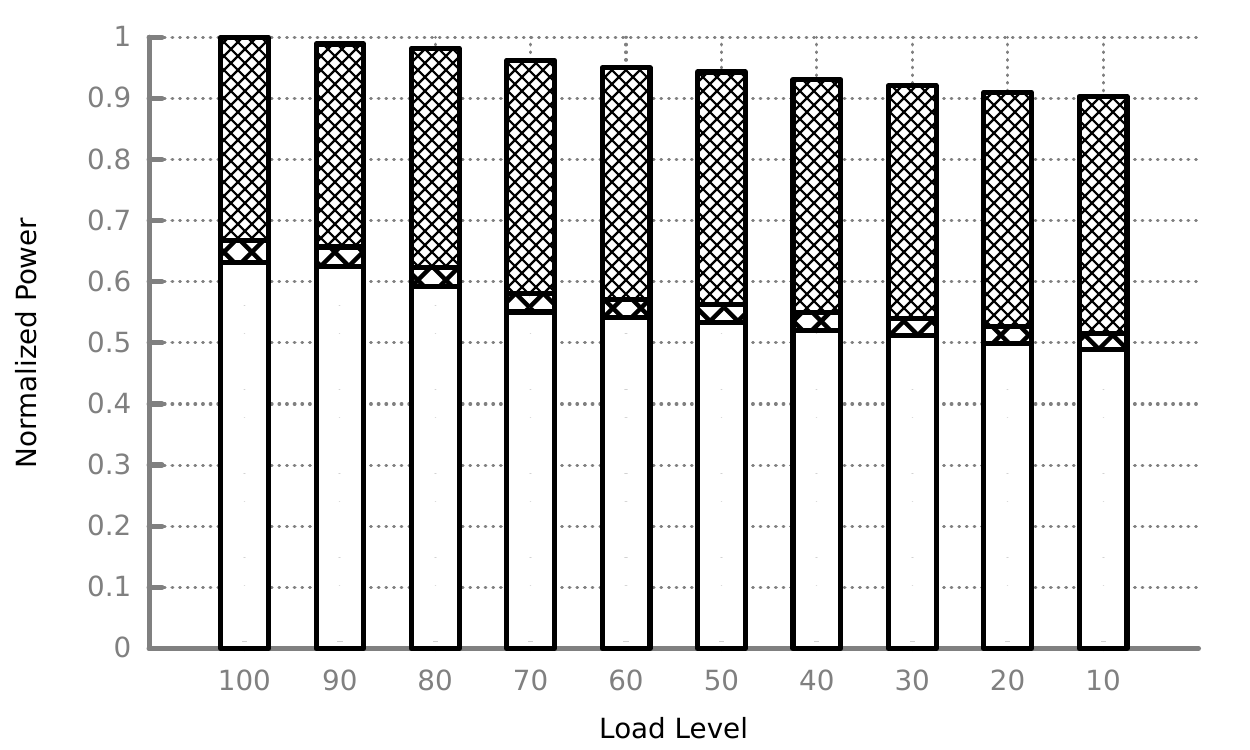}
\caption{Component-Level Power Distribution. Left: Data Serving (Cassandra), Center: 
Web Search (Nutch), Right: Data Caching (Memcached)}
\label{fig:pdis}
\end{figure*}

\emph{Idle Low-Power Modes:} These modes are applied when the processor is idling 
(i.e., no instruction is being executed). It is also referred as C-state. Each C-state 
selectively shuts down supporting circuitry in the processor in order to save more 
power. $CO$ state is the active state and $C1$ to $Cn$ (where, $n\in{1, 1E, 3, 6, 7}$) 
are idle low-power modes. $C7$ state saves the most power. Each mode has an exit latency 
(i.e., the time taken to bring back the processor to $C0$ state from any $Cn$ state) 
associated with it. Entering a deeper idle low-power mode incurs more cost in 
terms of exit latency as shown in Table~\ref{tab:cstatelat}.

\begin{table}[htb]
\centering
\caption{C-State Exit Latencies on Our Experimental Setup}
\label{tab:cstatelat}
\begin{tabular}{|c|c|c|c|c|c|} 
\hline
\textbf{C-State} & C1 & C1E & C3 & C6 & C7 \\ \hline
\textbf{Exit Latency (us)} &	2 & 10 & 80 & 104 & 109 \\ \hline
\end{tabular}
\end{table}  

To dynamically control C-states, we have to write the maximum allowable exit latency to the 
file \emph{/dev/cpu\_dma\_latency}. As long as the file \emph{/dev/cpu\_dma\_latency} is kept open, 
C-states with transition latencies higher than the specified exit latency value will not be used. 
For example, writing a maximum allowable latency of 12us will keep the processors only in C0, C1 or C1E 
state.

\subsubsection{Resource Provisioning}

Resource provisioning techniques aim to match the number of active servers 
to the load-level on the cluster. Idle servers can be simply turned off or 
provisioned to another application. However, there is an associated cost 
with bringing the servers back to active state as the load on the cluster increases. 
The evaluation of resource provisioning is done by manually hibernating nodes
in the cluster. In other words, we manually match the number of nodes to the load-level of the 
scale-out workload.

In the rest of the paper, we study the effect of these techniques on the 
energy proportionality and performance (in terms of response time/latency) 
of scale-out workloads.

\section{Component-Level Power Distribution}
\label{sec:powerdis}

In this section, we measure the component-level power consumption of 
the scale-out workloads under investigation. The main goal is to 
understand the contribution of each component to the energy 
proportionality of these workloads. 

Figure~\ref{fig:pdis} shows the power distribution
for the data serving, web search and data caching workloads for the entire 
cluster. The power consumption reported here corresponds to the processor P1 state with 
turbo and idle low-power modes disabled. In rest of the paper, this 
configuration is referred as no management (NM). The values reported in 
Figure~\ref{fig:pdis} are based on the sum of the power consumption from each node 
in the cluster and averaged over multiple runs. 
System components other than the processor and memory are represented
as \emph{Others}\footnote{The other components, denoted by ``Others,'' 
also include the power consumption of the hard disk.} in these figures. 

In the NM configuration, the power consumption of each component does not vary 
across load-levels. Such constant power consumption, irrespective of the load-level, 
is the main reason for poor energy proportionality of scale-out workloads. 
In general, the processor power consumption contributes most to the system power. 
The processor consumes 45-50\%, 65-70\% and 50-65\% 
for the data serving, web search and data caching workloads respectively. DRAM 
consumes lower power when compared to the processor across all the workloads. 
The power consumption of \emph{other} component varies depending upon the workload. 
However, their power consumption is always either similar (e.g., data serving) 
or less than the power consumption of the processor (e.g., web search). 

Since the processor consumes a large portion of the system power, this paper focuses on 
understanding the energy proportionality from the perspective of the processor. 
In rest of the paper, we study the utilization of the CPU and the effects of 
active low-power, turbo and idle low-power modes of the processor on the energy proportionality of the 
system for scale-out workloads. We also investigate the trade-offs involved in 
using resource provisioning and power provisioning.

\section{Analysis of CPU Utilization}
\label{sec:util}

Here we present the CPU utilization traces of the scale-out 
workloads for three load-levels at different time granularities. 
Using these results, we seek to understand whether power management 
techniques will have any effect on these workloads. Moreover, these 
experiments provide a view into the time granularity at which any power 
management technique should be applied. From the perspective of processor 
usage, these results also show the similarity and diversity of the workloads 
used in this paper. In this section, we will show the following: the potential 
to save power decreases with increase in time resolution and load-level and 
there is ample opportunity to save power at sub-second and sub-millisecond 
granularity at each load-level. 

\begin{figure*}[t]
\begin{minipage}[b]{1.0\linewidth}
\centering
\includegraphics[width=.329\linewidth,height=4cm]{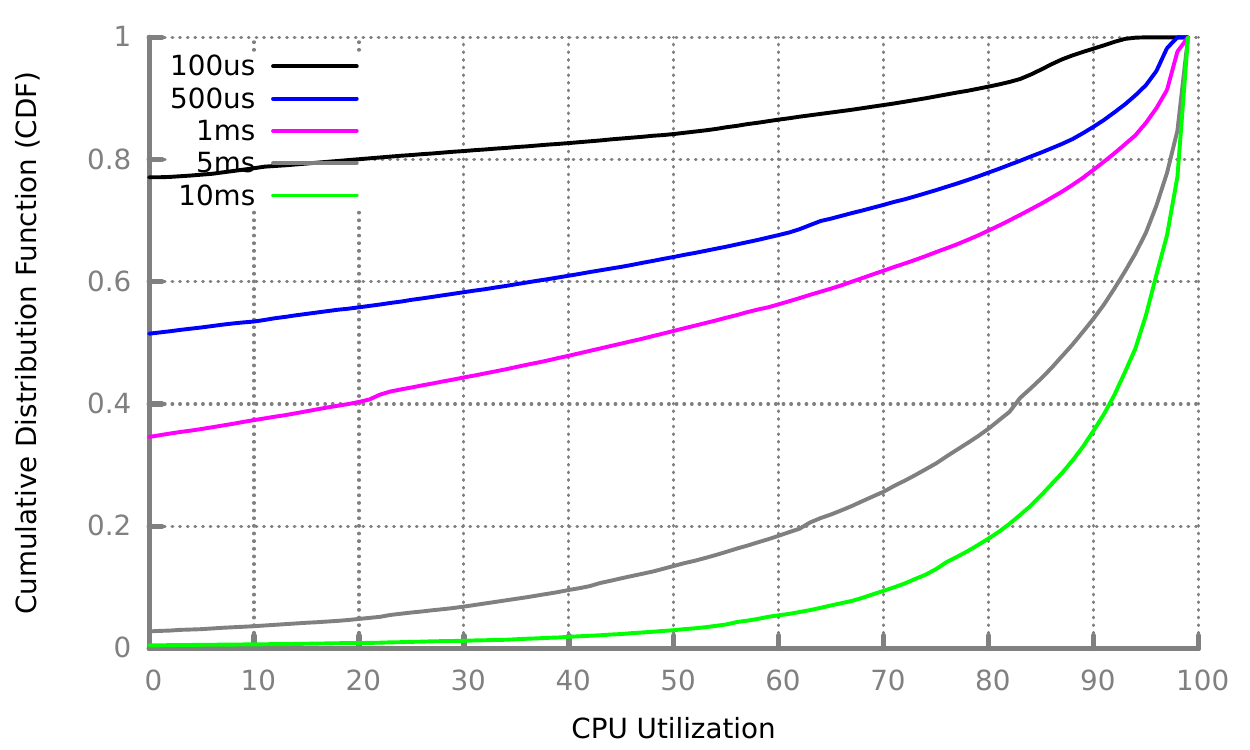}
\includegraphics[width=.329\linewidth,height=4cm]{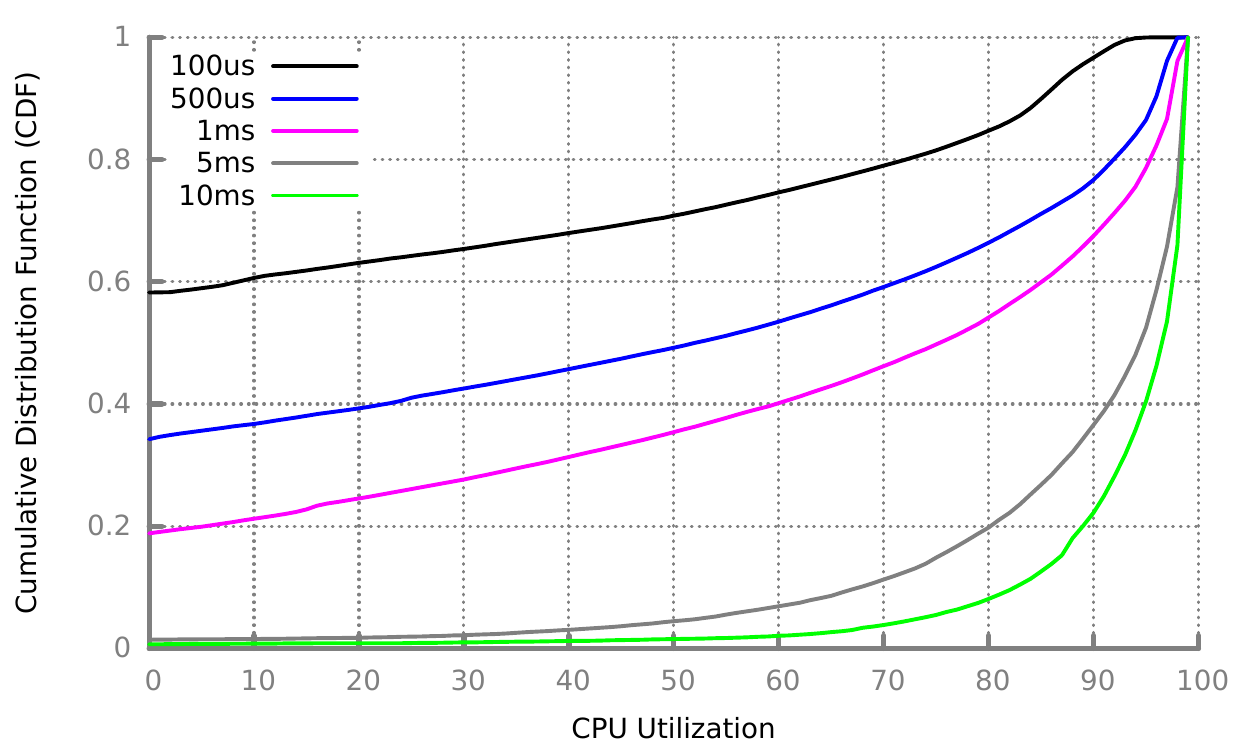}
\includegraphics[width=.329\linewidth,height=4cm]{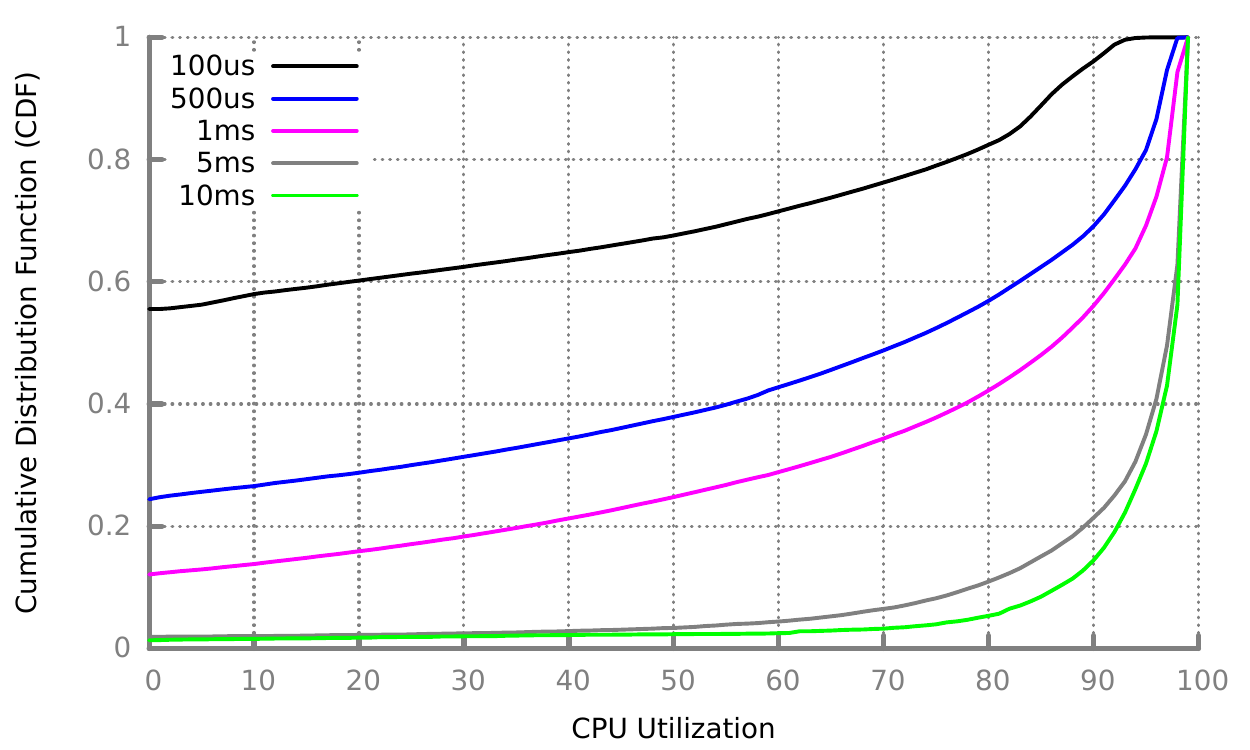}
\caption{CPU Utilization of Data Serving (Cassandra) At Different Time Resolution. 
30\% (Left), 50\% (Center) and 70\% (Right) Load-Levels}
\label{fig:utilac}
\end{minipage}

\begin{minipage}[b]{1.0\linewidth}
\centering
\includegraphics[width=.329\linewidth,height=4cm]{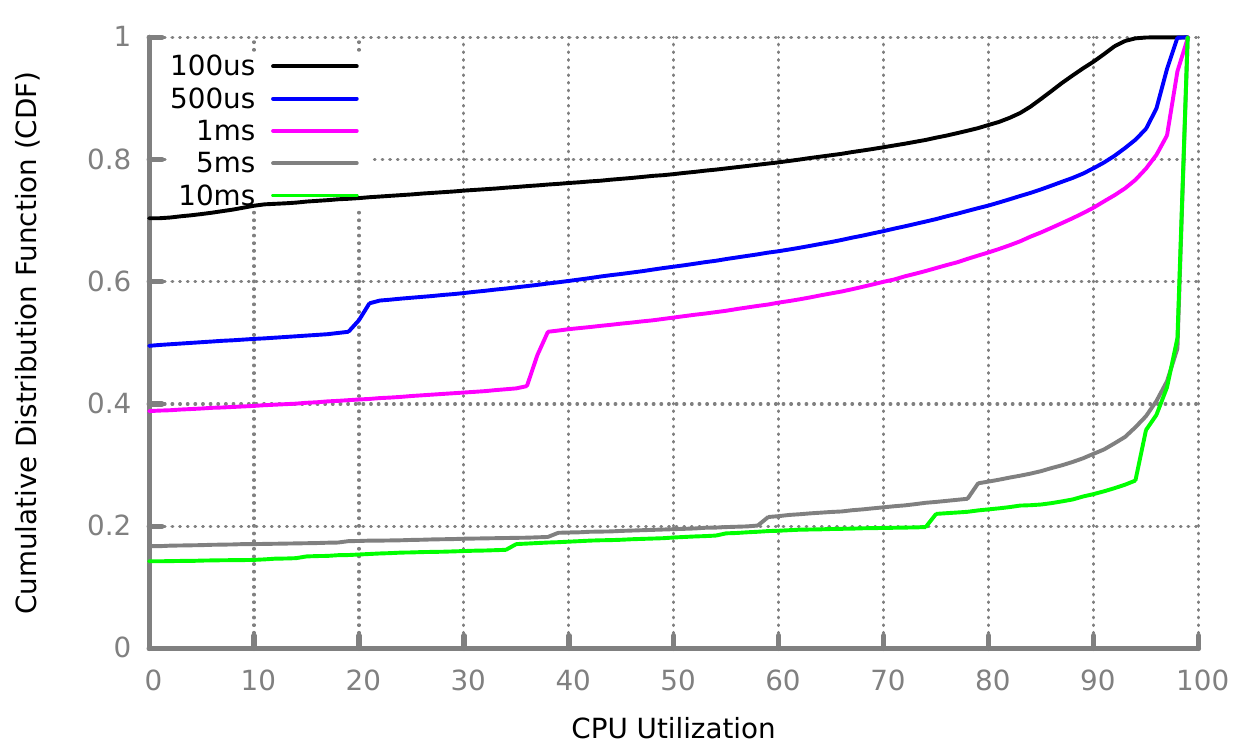}
\includegraphics[width=.329\linewidth,height=4cm]{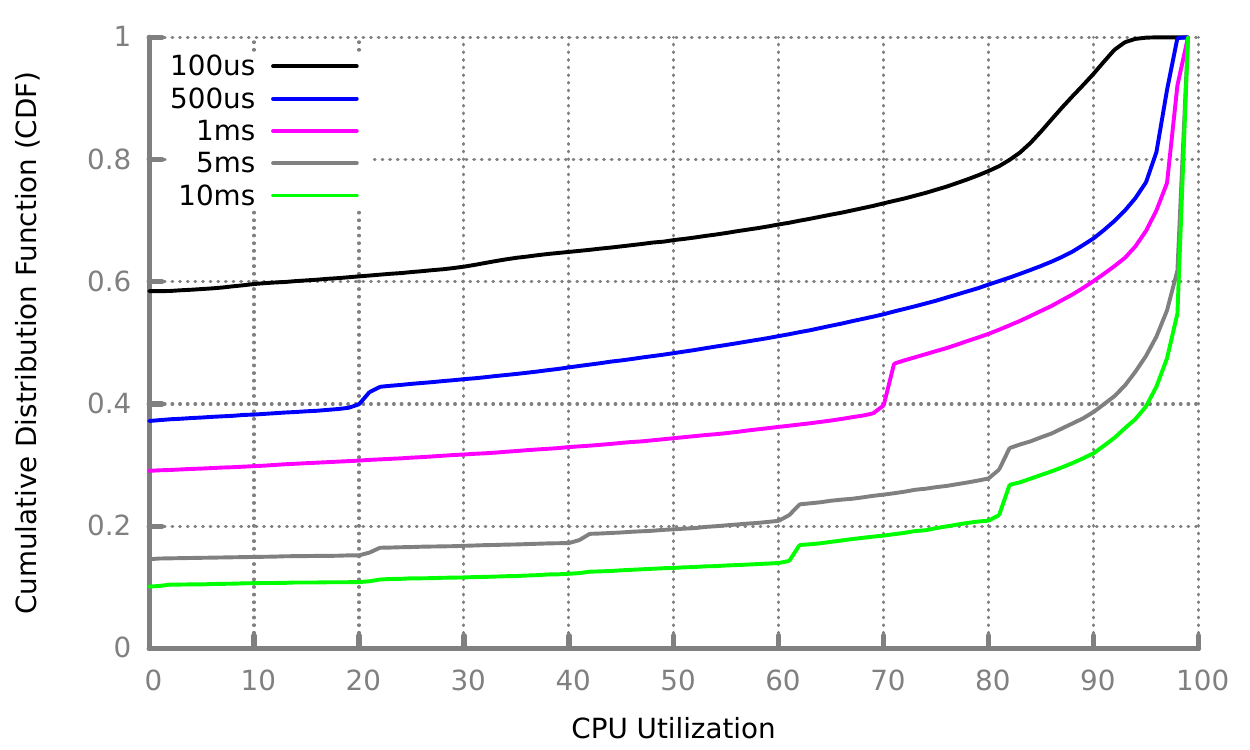}
\includegraphics[width=.329\linewidth,height=4cm]{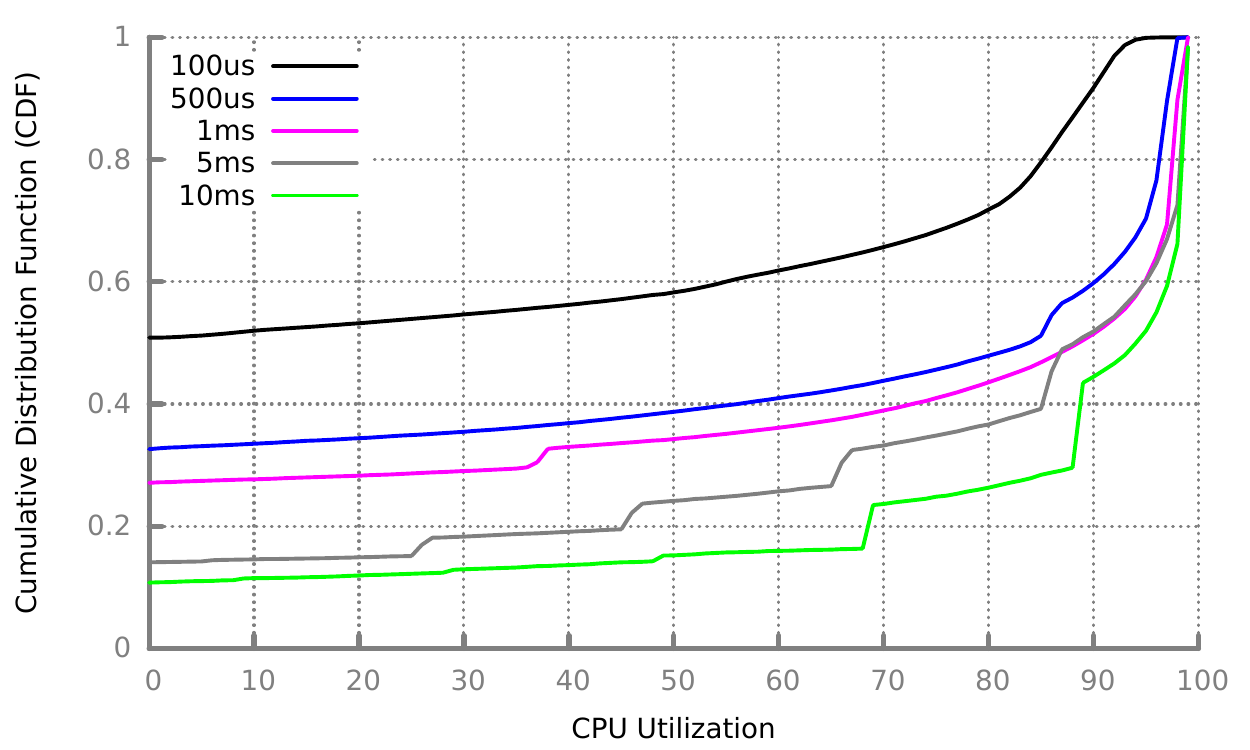}
\caption{CPU Utilization of Web Search (Nutch) At Different Time Resolution. 
30\% (Left), 50\% (Center) and 70\% (Right) Load-Levels}
\label{fig:utilws}
\end{minipage}

\begin{minipage}[b]{1.0\linewidth}
\centering
\includegraphics[width=.329\linewidth,height=4cm]{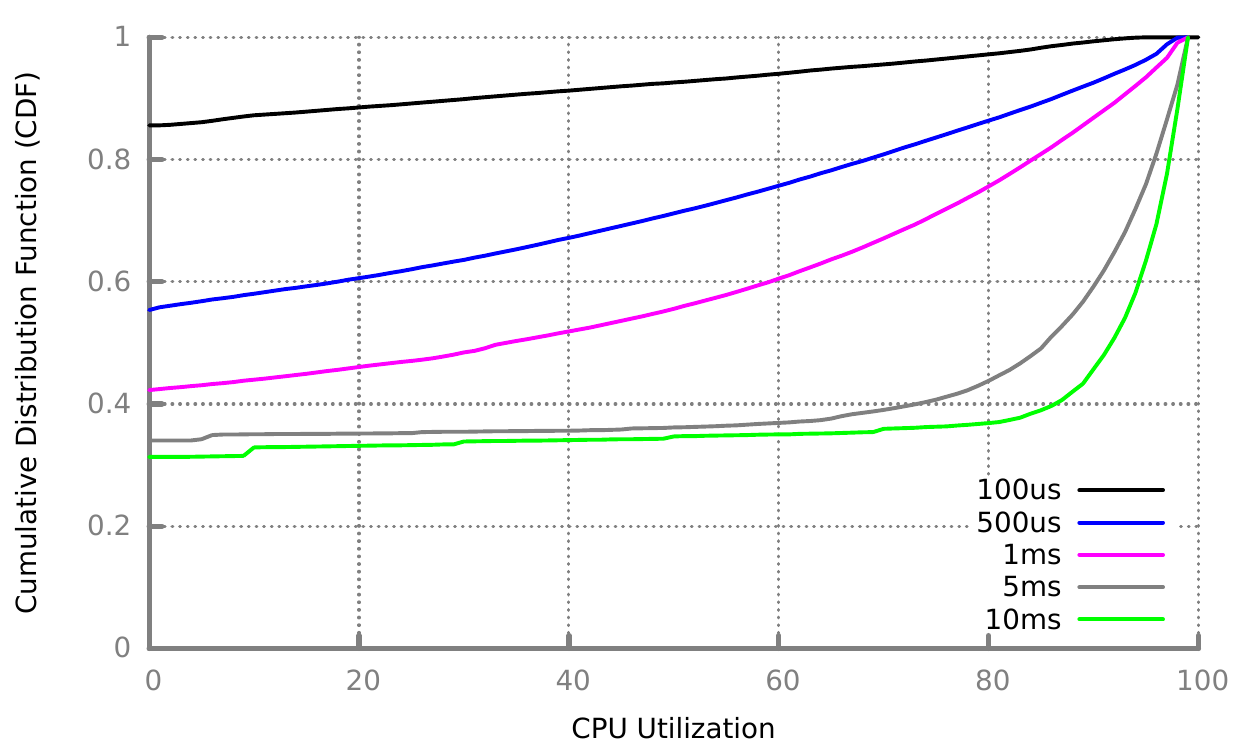}
\includegraphics[width=.329\linewidth,height=4cm]{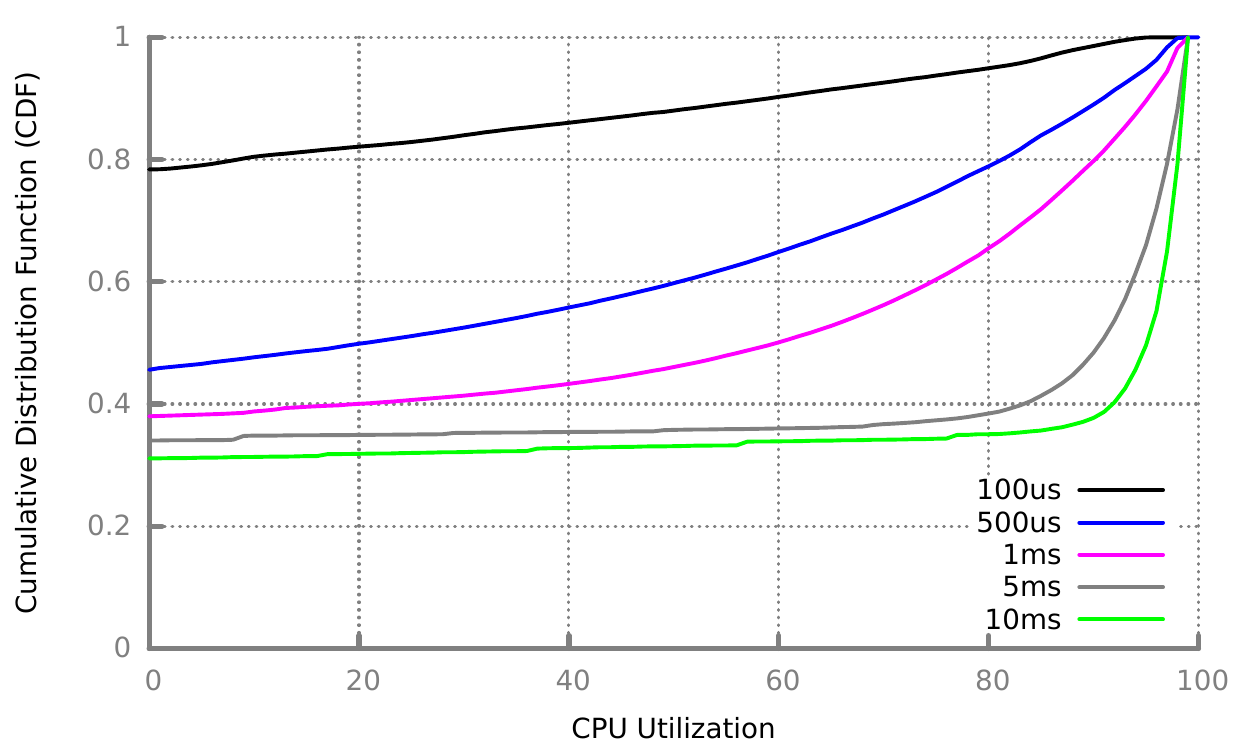}
\includegraphics[width=.329\linewidth,height=4cm]{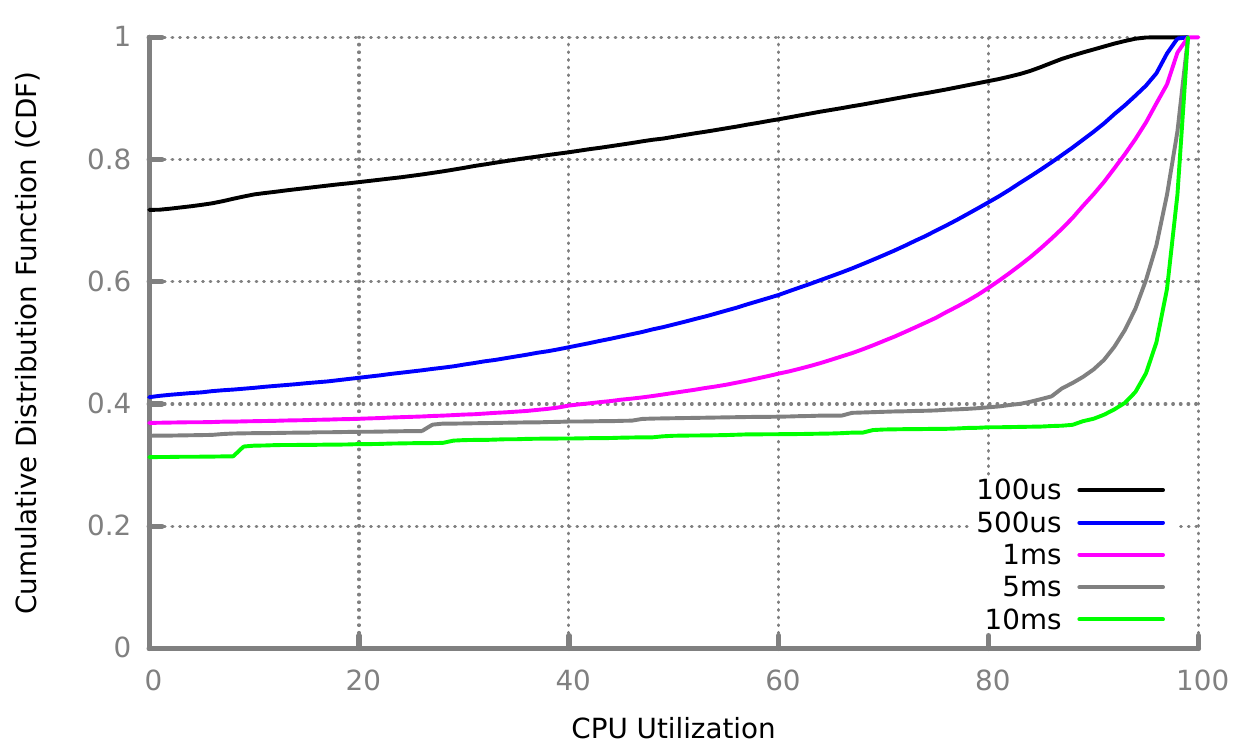}
\caption{CPU Utilization of Data Caching (Memcached) At Different Time Resolution. 
30\% (Left), 50\% (Center) and 70\% (Right) Load-Levels}
\label{fig:utilmc}
\end{minipage}
\end{figure*}

\subsection{Methodology}

The analysis of the CPU utilization is presented as cumulative 
distribution functions (CDFs). These CDFs provide a succinct view of the 
fraction of time for which a workload at a 
particular load-level spent its execution at or below a CPU utilization. 
We also show CDFs at  time granularities of 
100us, 500us, 1ms, 5ms and 10 ms. This allows us to visualize the change in 
CPU utilization as we change time resolution at which we monitor it. 
These fine-grained CPU utilization traces are collected using SystemTap~\cite{systemtap}.

\subsection{Discussion}

Figures~\ref{fig:utilac},~\ref{fig:utilws} and~\ref{fig:utilmc} show the 
CPU utilization profile of data serving, web search and data caching workloads, 
respectively, for three load-levels. Consider the 100us and 1ms CDFs 
for data serving workload (see 
Figure~\ref{fig:utilac}). The time spent at 0\% CPU utilization (i.e., idling) for both CDFs 
decrease when moving from 30\% load-level to 70\% load-level. It also 
decreases when you move from 100us to 1ms within a load-level. In general, 
the CDFs show that opportunity to save power and the amount of time spent 
idling decreases with increase in load-level and time resolution 
within a load-level. 

The fraction of time 
spent idling (i.e., 0\% utilization) at a time resolution (especially at 
sub-millisecond resolution) provides insights into the 
effectiveness of idle low-power modes on these workloads. At 5 and 10ms resolution, the data serving workload 
spends less time idling. Whereas, the other two workloads spend more time idling than the 
data serving workloads even at this resolution. It is expected 
that more time spent idling allows the processor to enter low power modes 
frequently and thus save power. These plots also provide insights into the time 
resolution at which to apply dynamic power management mechanisms. For examples, 
in all cases the best time resolution to apply power management mechanism is 100us 
as it provides best opportunity for power savings due to high fraction of idle time. 
Moreover, if the CDFs of two time resolution follow the similar curve 
it is better to  apply the power management mechanism  at the higher time 
resolution in order to avoid unnecessary overhead. For example, the 5 and 
10ms CDFs for 70\% load-level of the data serving and data caching workload follow a
similar curve. As a result, we can infer that there is no more opportunity to save 
processor power in 5ms than 10ms time resolution due to similar CPU utilization profile.

\section{Effect of Power Provisioning}
\label{sec:eprop}

\begin{figure*}[t!]
\begin{minipage}[b]{1.0\linewidth}
\centering
\includegraphics[width=.49\linewidth,height=6cm]{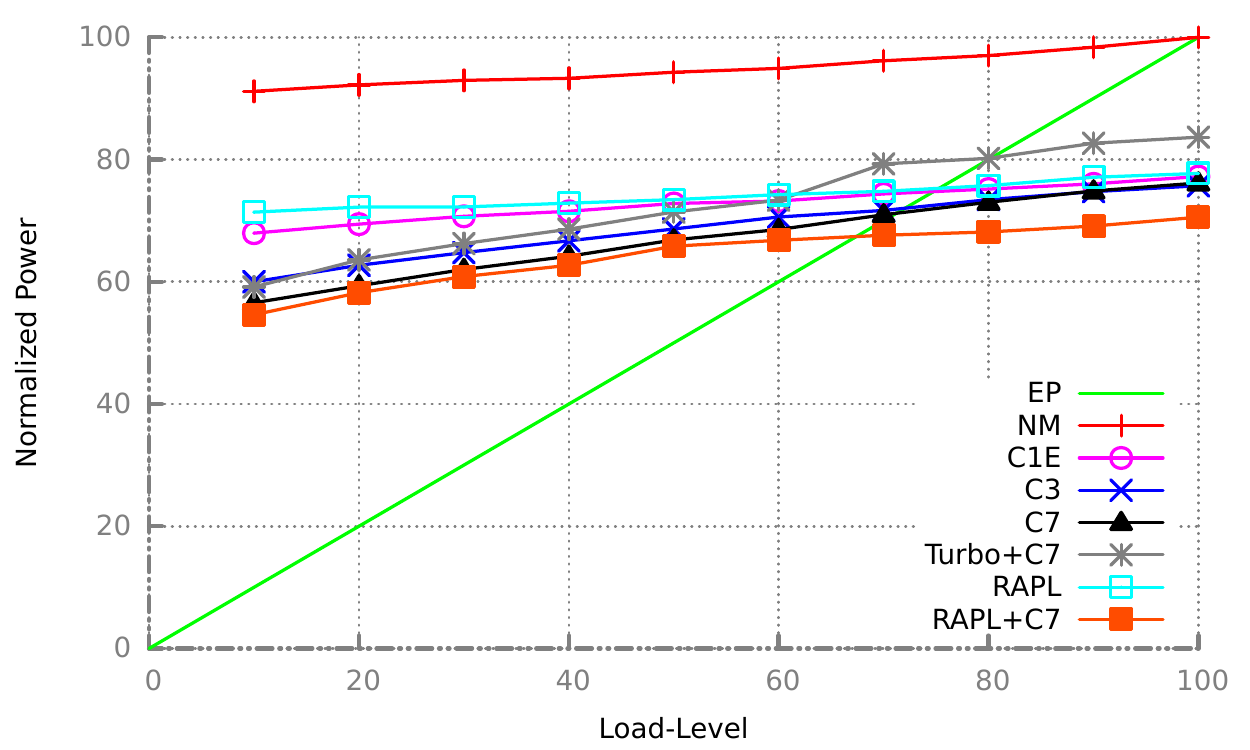}
\includegraphics[width=.49\linewidth,height=6cm]{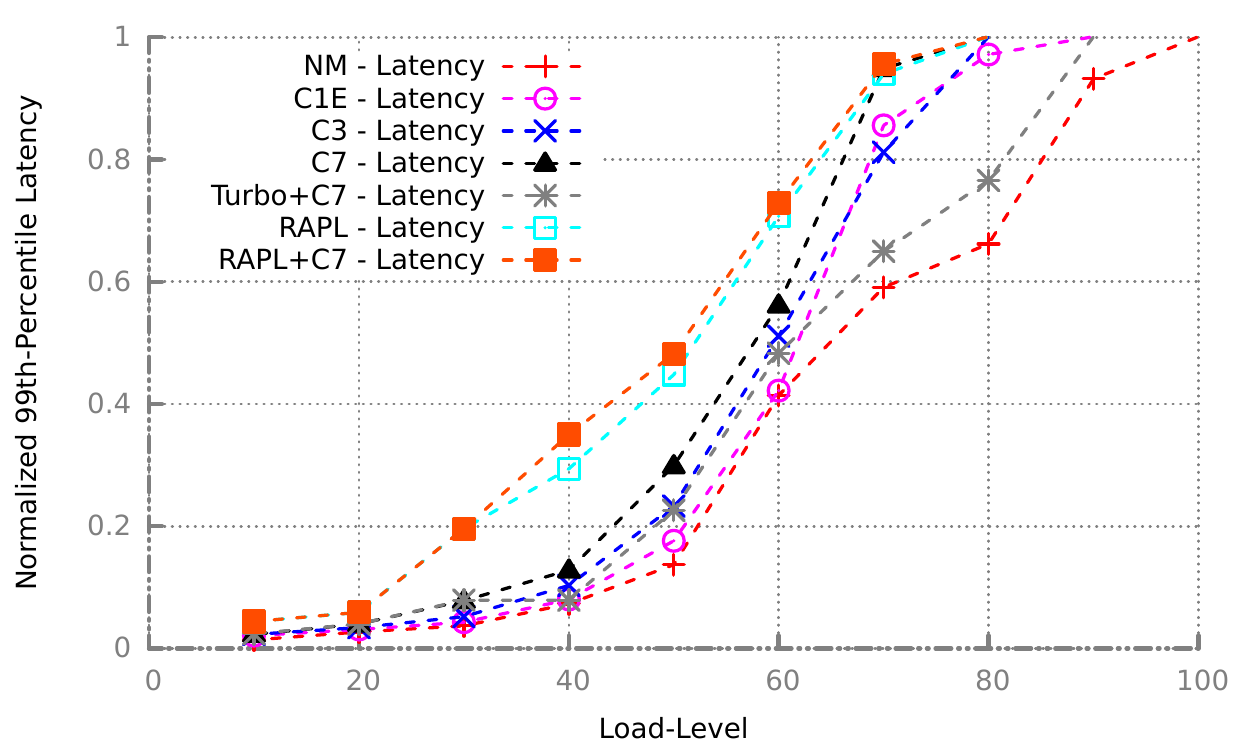}
\caption{Data Serving (Cassandra) Workload. 
Energy Proportionality (Left) and Latency (Right). NM = No Management.}
\label{fig:epac}
\end{minipage}

\begin{minipage}[b]{1.0\linewidth}
\centering
\includegraphics[width=.49\linewidth,height=6cm]{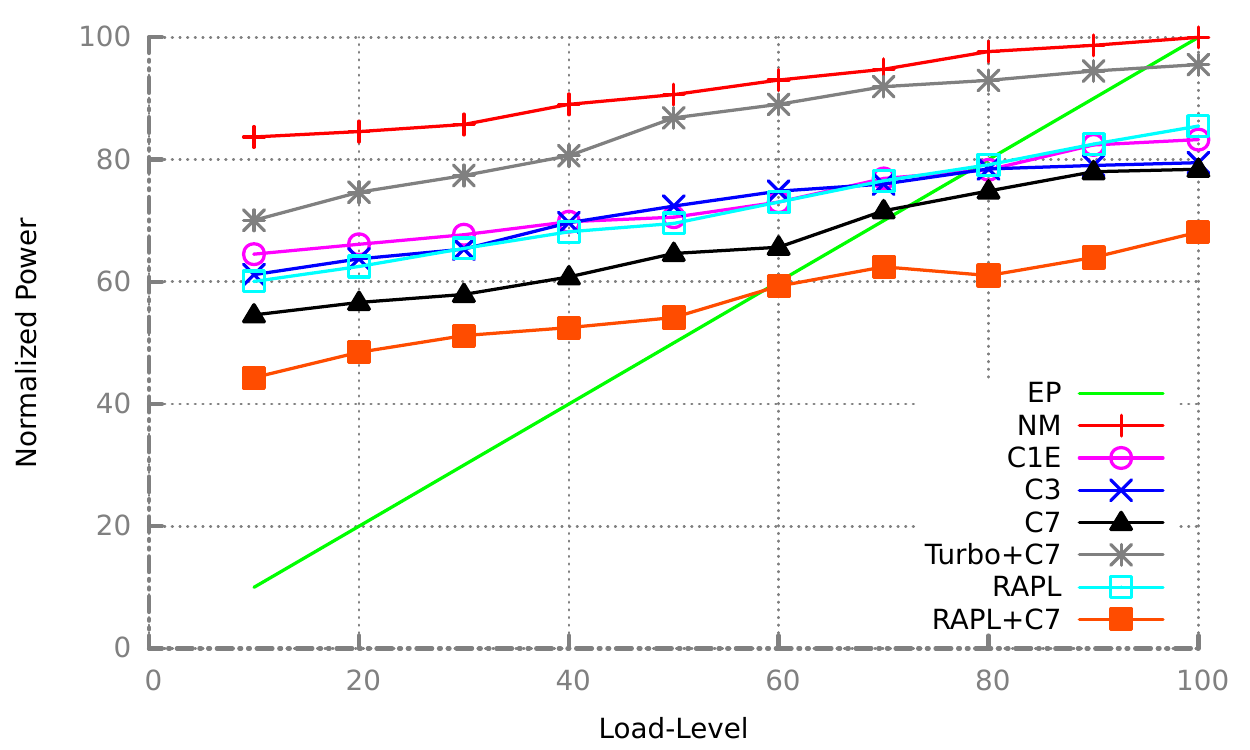}
\includegraphics[width=.49\linewidth,height=6cm]{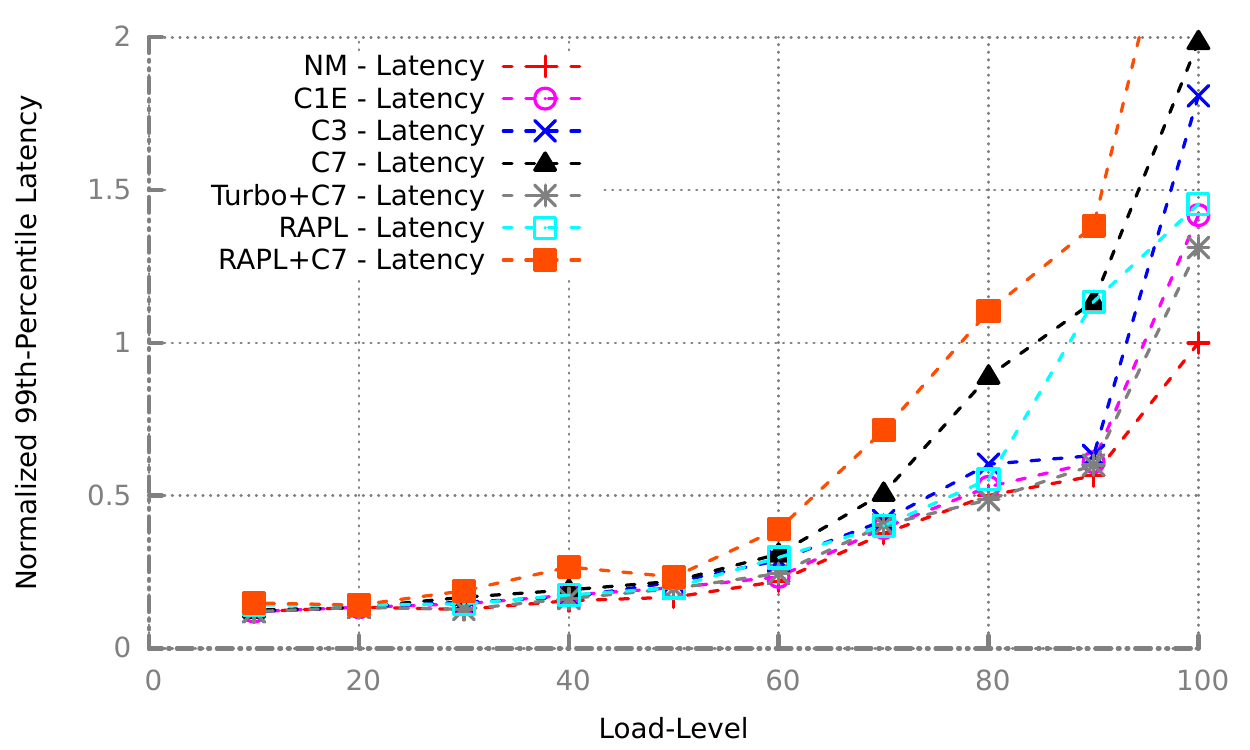}
\caption{Web Search (Nutch) Workload. 
Energy Proportionality (Left) and Latency (Right). NM = No Management.}
\label{fig:epws}
\end{minipage}

\begin{minipage}[b]{1.0\linewidth}
\centering
\includegraphics[width=.49\linewidth,height=6cm]{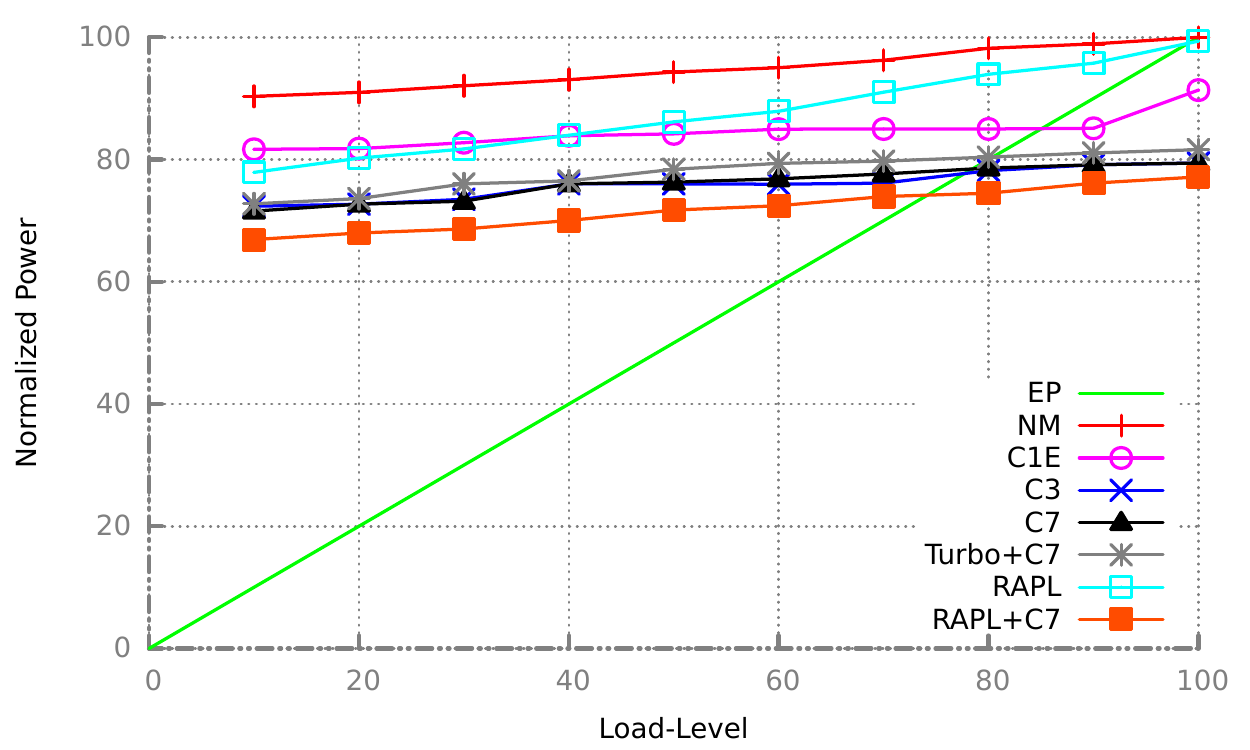}
\includegraphics[width=.49\linewidth,height=6cm]{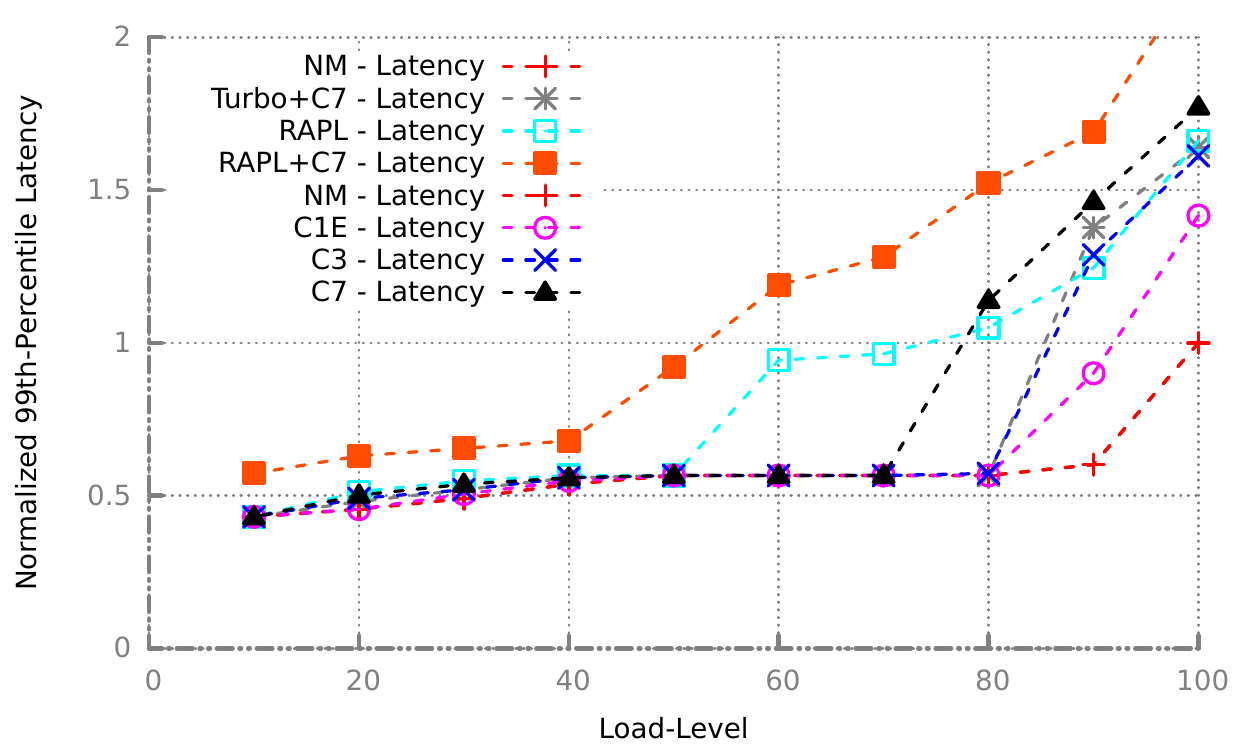}
\caption{Data Caching (Memcached) Workload. 
Energy Proportionality (Left) and Latency (Right). NM = No Management.}
\label{fig:epmc}
\end{minipage}
\end{figure*}

In this section, we describe the effects of the power provisioning 
techniques (i.e., active low-power, turbo and idle low-power modes) on the 
energy proportionality, corresponding response times (latency) and power saving 
potential for the scale-out workloads. 

\begin{figure*}[t]
\centering
\includegraphics[width=.329\linewidth,height=4cm]{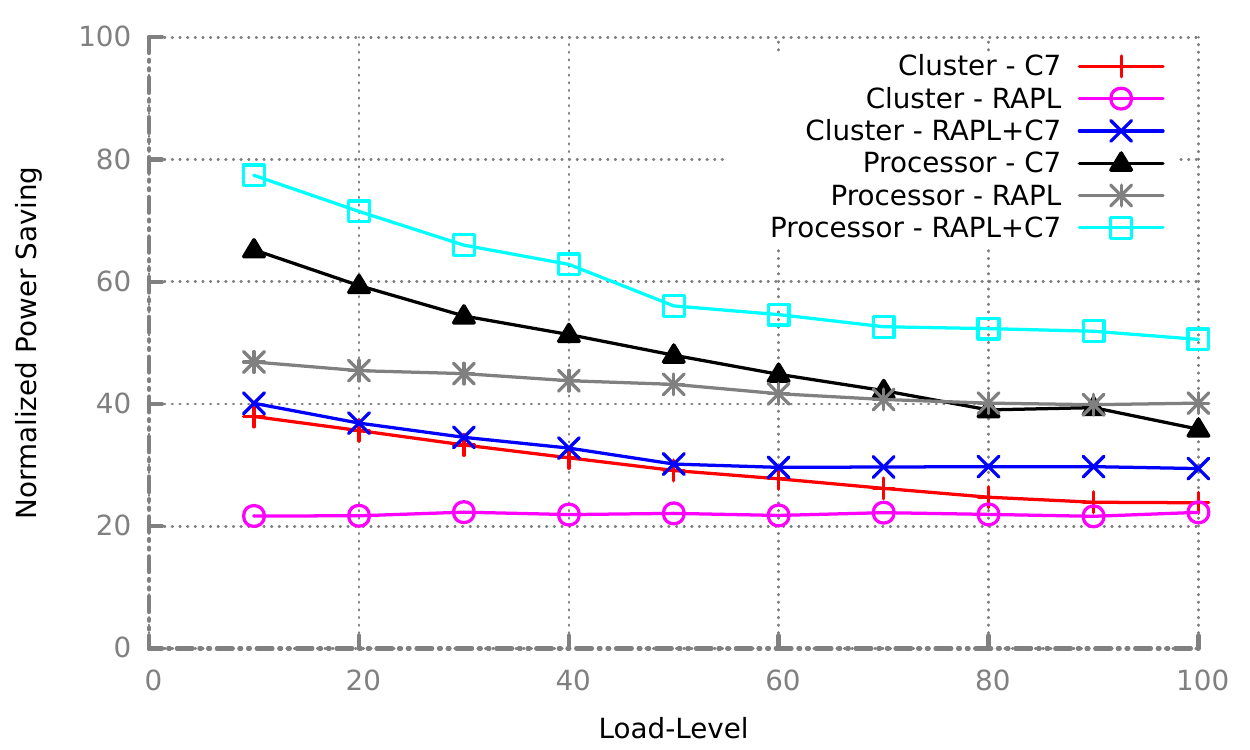}
\includegraphics[width=.329\linewidth,height=4cm]{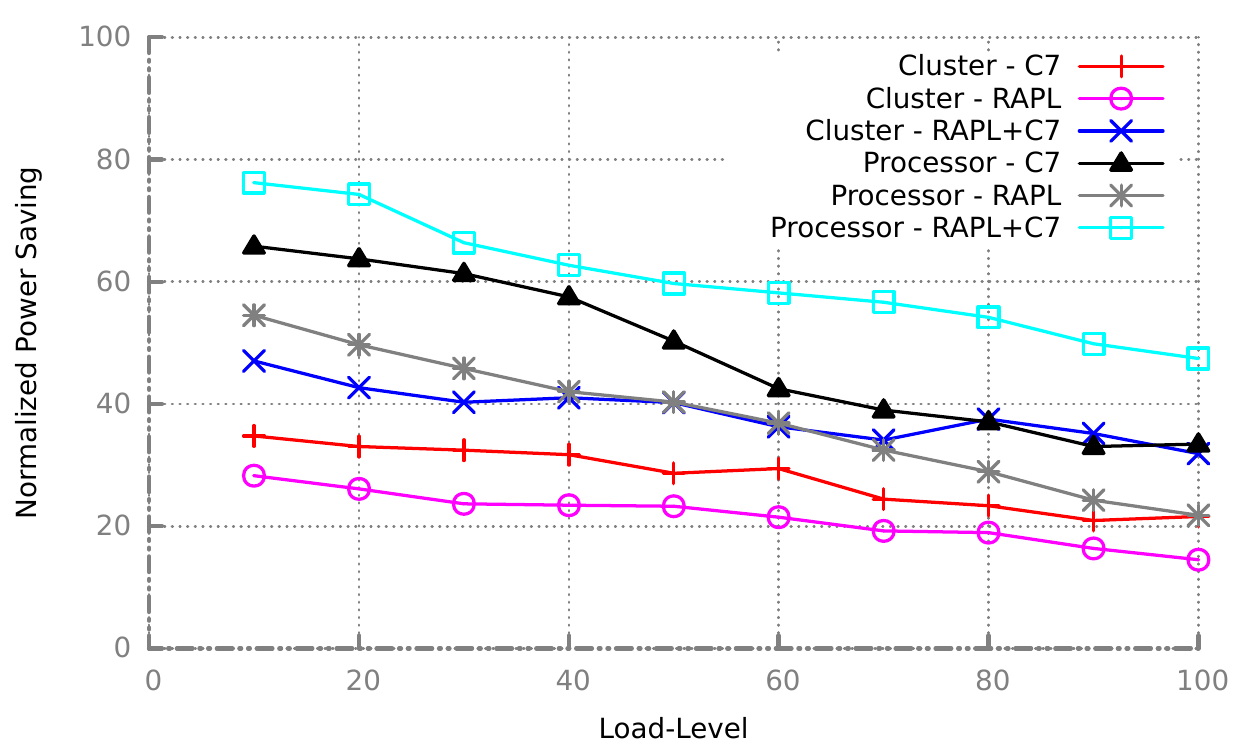}
\includegraphics[width=.329\linewidth,height=4cm]{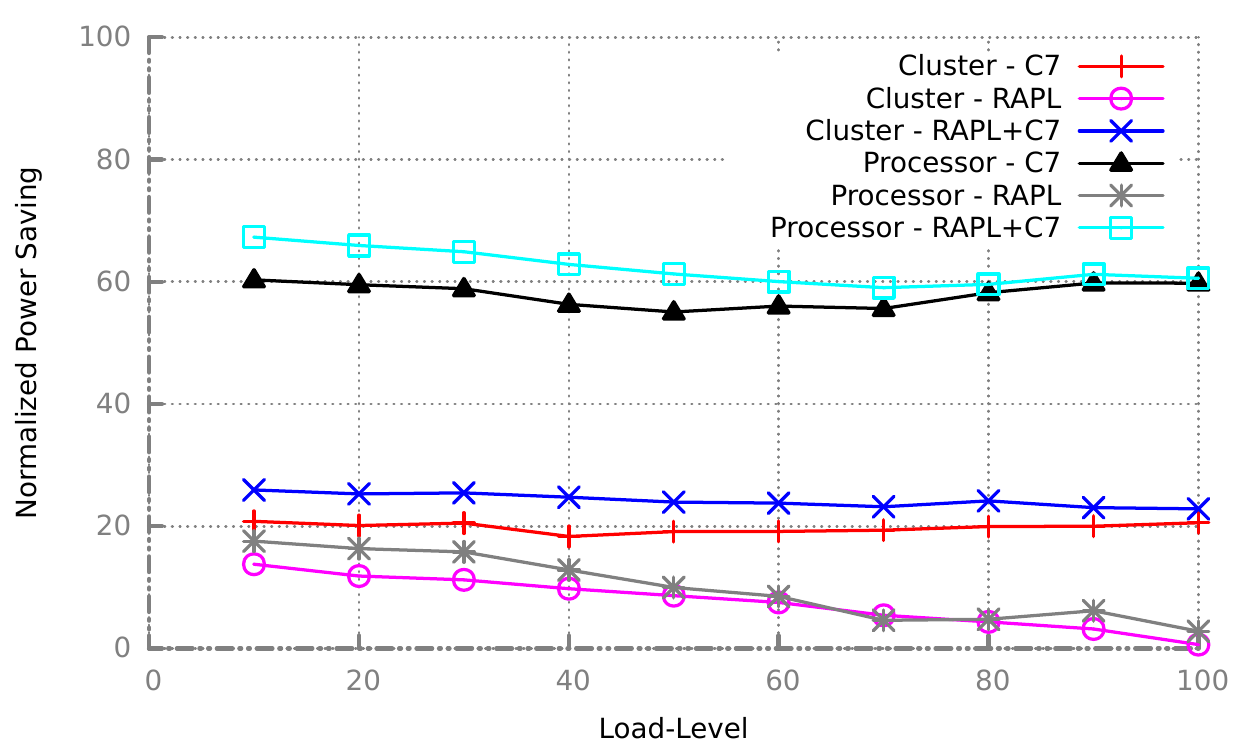}
\caption{Power Savings. 
Data Serving (Left), Web Search (Center) and Data Caching (Right).}
\label{fig:psavings}
\end{figure*}

We describe these effects using Figures~\ref{fig:epac}, \ref{fig:epws}, \ref{fig:epmc} and~\ref{fig:psavings}. 
The No Management (denoted as NM) line in each figure corresponds to the power consumed by each 
workload when the system is restricted to P1 state (i.e., 2.0GHz) and C0 state with 
turbo mode disabled. For studying the effects of idle low-power modes, we restrict the 
deepest possible C-state that can be entered by writing appropriate latency values 
to the \emph{/dev/cpu\_dma\_latency} device file as described in Section~\ref{sec:bkgndpprov}. 
The effects of turbo mode on these workloads are studied using the procedure 
described in Section~\ref{sec:bkgndpprov}. To understand the potential of active low-power 
modes, we run through all possible RAPL processor power limits possible on each load-level  
such that the throughput is maintained. In this section, 
we present only the results from the best possible RAPL configuration (i.e., the 
configuration which achieves the best power savings while meeting throughput 
constraints). We also show the effects of RAPL in conjunction C7 state (denoted as RAPL+C7). 
We do not show results which restricts the deepest possible C-state to C1 and C6 
because C1E and C7 had identical effects to C1 and C6, respectively, for all the workloads. 
In all cases, using only turbo mode resulted in power consumption that exceeded the power 
consumption of each workload at 100\% load-level. Hence, we do not show results corresponding 
to only turbo mode. The results corresponding to turbo mode in conjunction with C7 state 
(denoted as Turbo+C7) is shown instead. 

\subsection{Effects on Energy Proportionality}

The left plots in Figures~\ref{fig:epac}, \ref{fig:epws} and~\ref{fig:epmc} 
show the effect of power provisioning on the energy proportionality of scale-out 
workloads. The Y-axis represents the normalized power (normalized to the power consumed at 100\% 
load-level) by the system and X-axis represents the load-level. As a result, the energy-proportional curve 
(denoted by EP) consumes 40\% of power at 40\% load-level, 60\% of power
at 60\% load-level and so on.

As observed, better than energy-proportional operation is not possible under all load-levels for these 
workloads. However, energy-proportional operation can be achieved for certain load-levels depending 
upon the workload. For example, better than energy-proportional operation can be achieved for 
load-levels greater than 70\%, 60\% and 80\% load-level for data serving, web search and data caching workloads, 
respectively, using RAPL+C7. But energy proportionality is improved in every case. 

\begin{figure}[h]
\centering
\includegraphics[width=1.0\columnwidth]{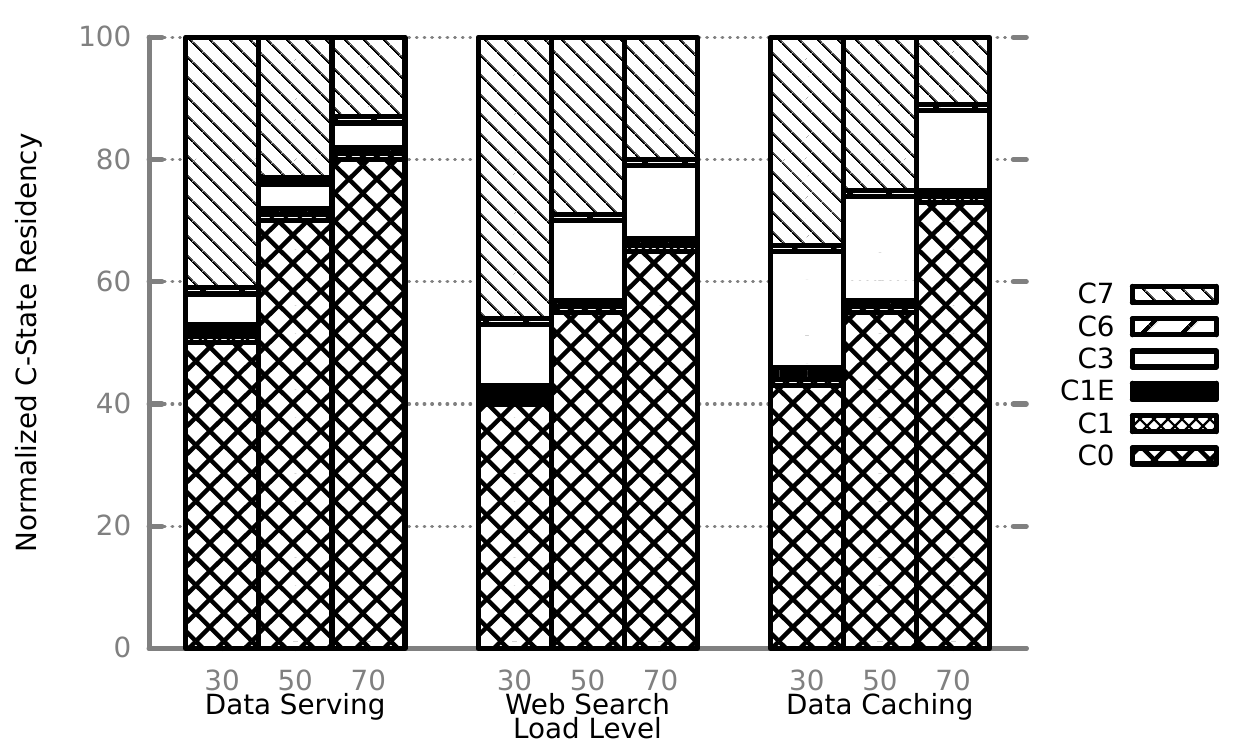}
\caption{C-State Residency of Scale-Out Workloads}
\label{fig:cstateall}
\end{figure}

Overall, RAPL+C7 achieves the best energy proportionality. But only in certain cases, C7 
state improves energy proportionality as much as RAPL+C7. However, in Section~\ref{sec:util}, 
we showed that the there is ample idle time available for the idle low-power modes to save power 
even at sub-millisecond time granularity. To understand the reasons behind C7 state not performing as 
expected, we looked at the C-state residency (i.e., time spent in each C-state) shown in 
Figure~\ref{fig:cstateall}. As observed, the processor does not spend a amount of time 
proportional to idle time in deep idle low-power states such as C3, C6 or C7 even though there 
is ample opportunity to do so in the 100us and 500us time granularities. A huge fraction of 
time is spent in C0 state even though the CPU is idling. From these results, 
we infer that there is a need for better system software to enter deep low-power states more 
aggressively in order to save power. 

However, the best idle low-power mode 
power savings come from allowing the processor to use the C7 state as expected. Using, turbo+C7 
does not improve energy proportionality for high load-levels. But some improvement is seen when turbo+C7 
is used at low load-levels. However, it does not do better than using only C7. While these plots only 
describe the improvements in energy proportionality, the scale-out workloads are latency-sensitive. 
The latency of these workloads might be affected due to power management techniques even though 
the throughput (i.e., load-level) can be maintained. We discuss the effects of using these power 
management techniques on the latency of scale-out workloads in the next section.

\subsection{Effects on Latency (Response Time)}

The plots in the right of Figures~\ref{fig:epac}, \ref{fig:epws} and~\ref{fig:epmc} 
show the effect of power provisioning on the 99th-percentile latency (i.e., response times) 
of scale-out workloads. These curves represents a typical response time curve for a scale-out workload.
Each curve has a inflection point after which the response times 
rapidly rises. This inflection point for NM curve lies at 50, 60 and 90 percent load-level for the 
data serving, web search and data caching workloads, respectively. It is clear from the plots 
that each power management technique has its own unique effect on the response times of the 
workload. For example, the inflection point is shifted when a more aggressive power management 
technique is used. We also observe that for certain load-levels all the latency curves are 
overlapping. In case of web search, all the latency curves overlap after 60\% load-level 
suggesting that a more aggressive power management technique can be used without having 
any effect on the response times of this workload at or below 60\% load-level. Similar 
inferences can be made for other workloads and power management techniques. 

In practice, these workloads operate under strict service-level objectives (SLOs). This SLO is placed
on the 99th-percentile latency and fixed for a particular workload. Any violation of SLO will be 
unacceptable. As such, a mechanism to trade-off some power savings for improvements in response times is 
desirable. In Section~\ref{sec:pvlat}, we describe the potential of RAPL to provide a trade-off 
between power and latency. This makes dynamic power management using active low-power modes 
an attractive option for improving the energy proportionality of scale-out workloads while meeting 
strict SLOs.

\subsection{Power Savings}

Figure~\ref{fig:psavings} shows the power saving achieved for different load-levels of the scale-out workloads. 
We show the power savings at the cluster- and processor-level. The processor-level 
power savings include the combined power consumption of all the processors in the 
cluster. This processor power consumption is measured using RAPL interface. We only show 
the processor-level power consumption as the turbo, active low-power and idle low-power 
modes affect only the processor power consumption. The power 
saving is calculated as $(power_{NM} - power_{mode})/power_{NM}$. At the 
cluster-level, we save the highest power using RAPL+C7 in all 
cases. The difference in power savings achieved due to RAPL+C7 when compared to only C7 
is substantial for each workload. We save up to 40, 47 and 26 percent 
for the three different workloads at the cluster-level when compared to the 
power consumption with no management. While comparing the power saving at 
the processor-level, the difference between the power savings achieved 
with RAPL+C7 and C7 is even more pronounced. The maximum power savings 
achieved for the scale-out workloads at the processor-level are 77, 76 and 67 
percent.

\section{Trade-Offs in Power and Latency}
\label{sec:pvlat}

\begin{figure*}[t]
\centering
\includegraphics[width=.329\linewidth,height=4cm]{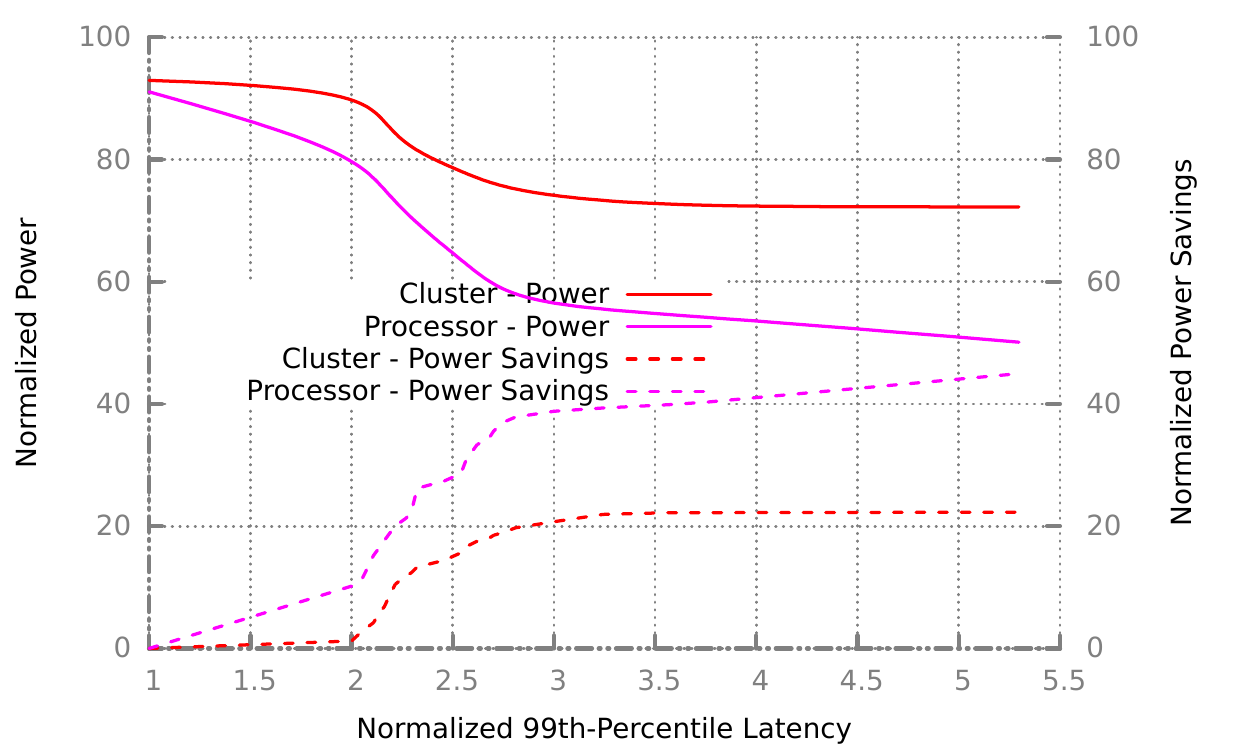}
\includegraphics[width=.329\linewidth,height=4cm]{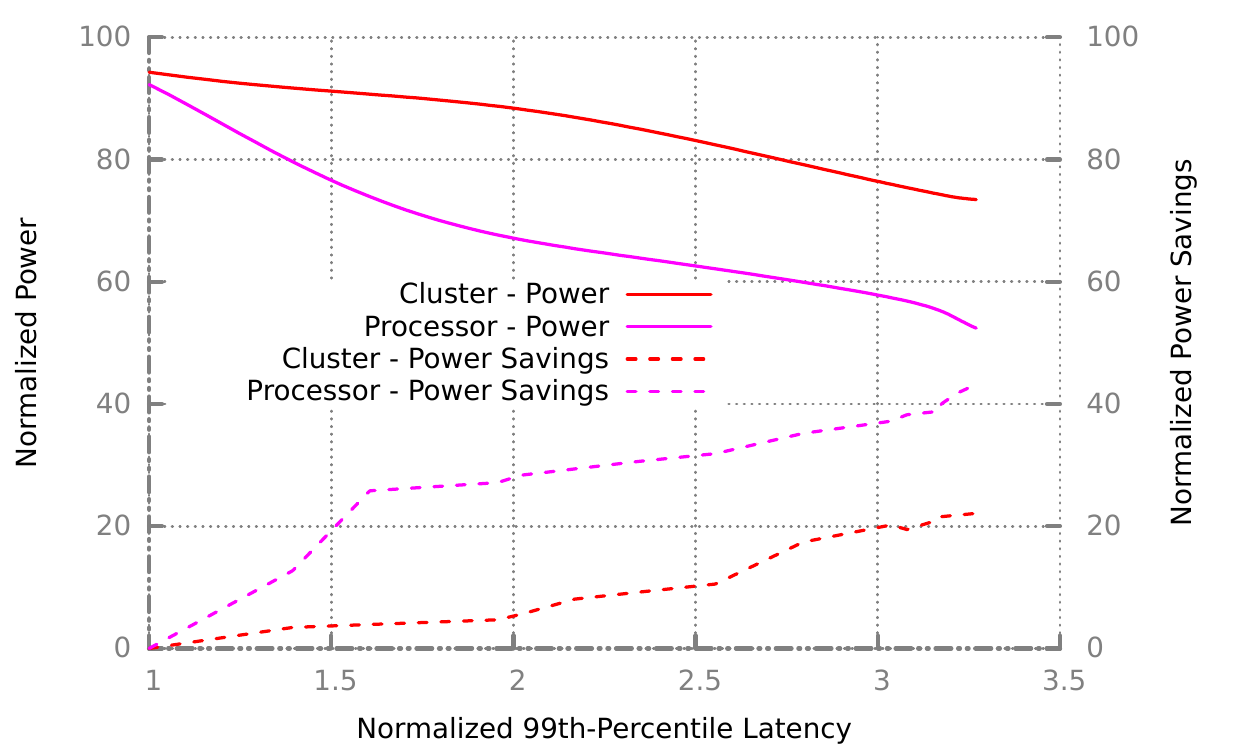}
\includegraphics[width=.329\linewidth,height=4cm]{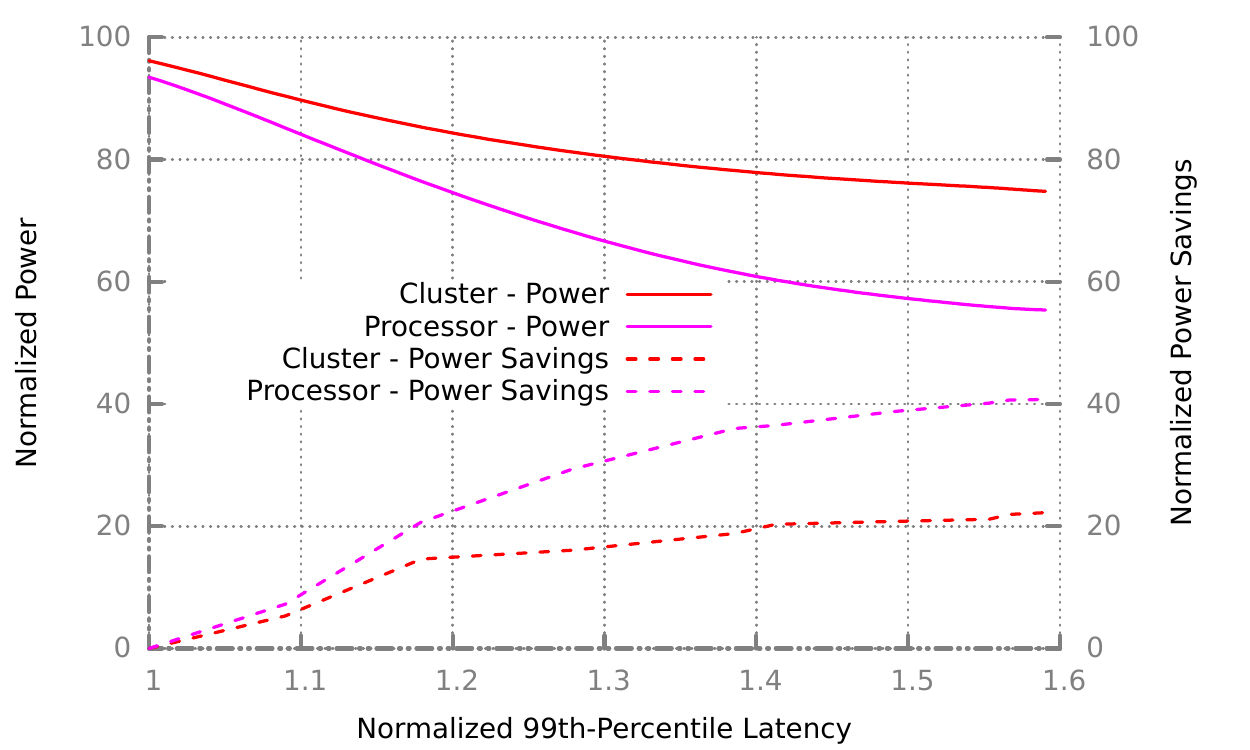}
\caption{Trade-Offs in Power and Latency. Left: 30\%, Center: 50\%, 
Right: 70\% Load-Levels) of Data Serving Workload}
\label{fig:acpvlat}
\centering
\includegraphics[width=.329\linewidth,height=4cm]{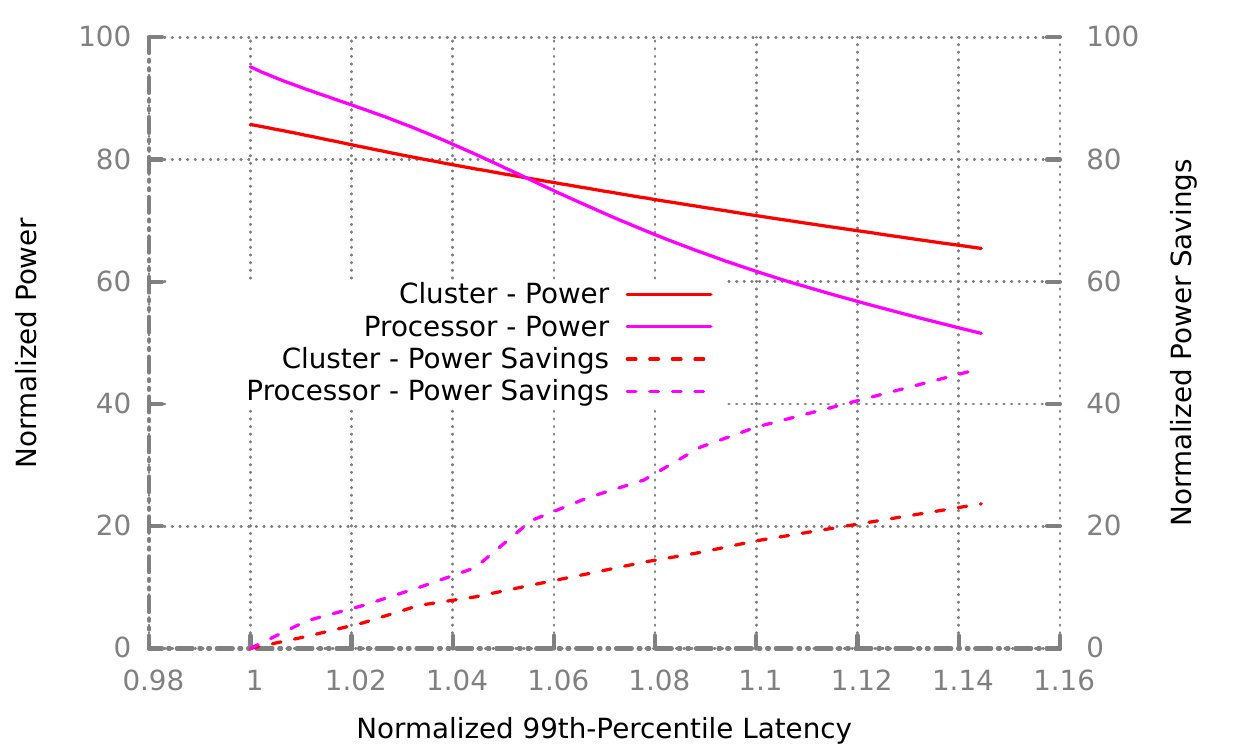}
\includegraphics[width=.329\linewidth,height=4cm]{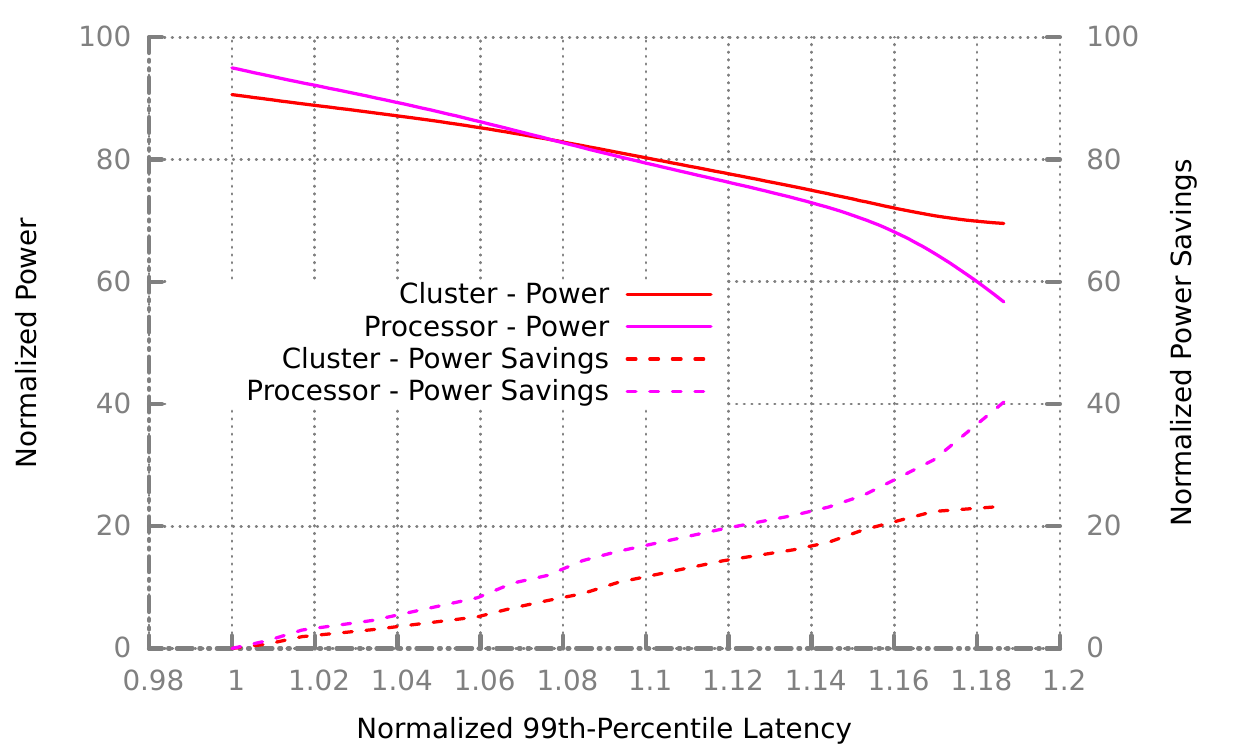}
\includegraphics[width=.329\linewidth,height=4cm]{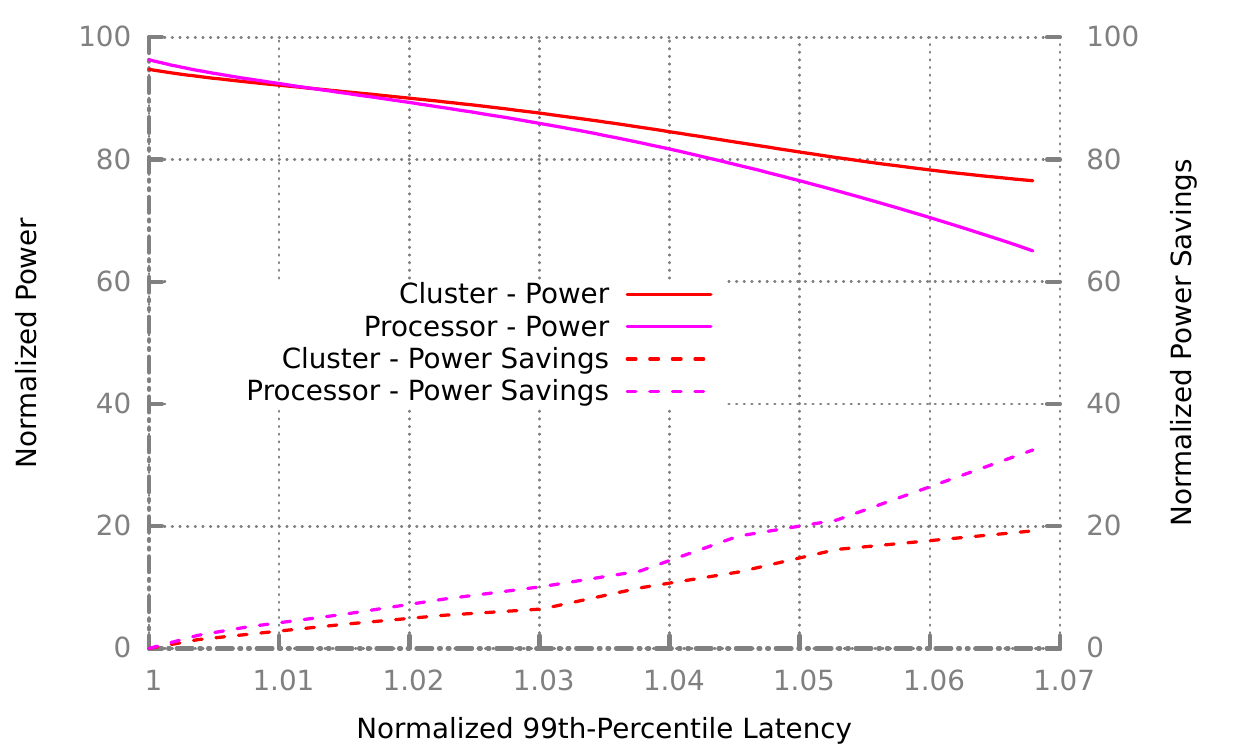}
\caption{Trade-Offs in Power and Latency. Left: 30\%, Center: 50\%, 
Right: 70\% Load-Levels) of Web Search Workload}
\label{fig:wspvlat}
\end{figure*}

As mentioned earlier, active low-power modes not only provide the 
best power saving but also gives us an opportunity to operate a given 
workload in a configuration to meet strict SLOs by creating a 
power-performance trade-off space. In this section, we describe this 
space with respect to scale-out workloads. Our methodology involves the 
projection of power consumed for a particular latency while running 
the scale-out workload on our experimental testbed. We run through 
different possible RAPL processor power limits and present the impact 
of each configuration on the power and performance (latency) of these workloads.

Figures~\ref{fig:acpvlat} and~\ref{fig:wspvlat} show the power-performance trade-off space for 
the 30, 50 and 70 percent load-level of the data serving and web search workload, respectively.
In this figures turbo mode was disabled and the processor was restricted to C0 state. Essentially, it is 
power-performance exhibited by only using RAPL.\footnote{We do not show results for the data caching 
workload since it does not exhibit any variation for latency in RAPL mode for 30\% and 50\% load-level 
as shown in Figure~\ref{fig:epmc}.} The X-axis represents normalized 99th-percentile latency. 
The Y-axes represents normalized power and power savings, respectively. 
The latency and power saving are normalized 
w.r.t. no management (NM) mode at same load-level. However, power is normalized to the 
power consumed at 100\% load-level in order to illustrate energy proportionality. 
For 70 and 50 percent load-levels of the data serving workload, the power vs. latency 
curve follows a similar trend at cluster-level. They provide best power savings on the left side of the graph 
where the latency penalty is less. However, as more aggressive power limiting is applied, the 
power savings achieved stagnates. In case of the 30\% load-level at the cluster-level, in the initial run through of the 
power limits, the latency increase more that the power savings it provides, in the center part of the plot, 
there is an inflection point which provides best power savings for little increase in latency and finally, 
the power savings achieved stagnates. For all load-levels at the processor-level, the power vs. latency curve 
follows a similar trend. Similar observations can be made for the power-performance trade-offs 
of the web search workload as shown in Figure~\ref{fig:wspvlat}. 
Moreover, the trade-off space exhibited by each workload depends upon the load-level. For example, the web search 
workload gives an opportunity to operate it anywhere in the $1-1.15x$ latency cost 
range depending upon the load-level. 
Such inferences on the relationships between power and latency are useful to design a 
dynamic power management runtime for scale-out workloads. 

\section{Power vs. Resource Provisioning}
\label{sec:ppvrp}

In this section, we evaluate the impact of resource and power provisioning 
techniques on the energy proportionality and power savings achieved. Our main goal is 
to compare and contrast the above two in the context of scale-out workloads. 

\begin{figure*}[t]
\centering
\includegraphics[width=.329\linewidth,height=4cm]{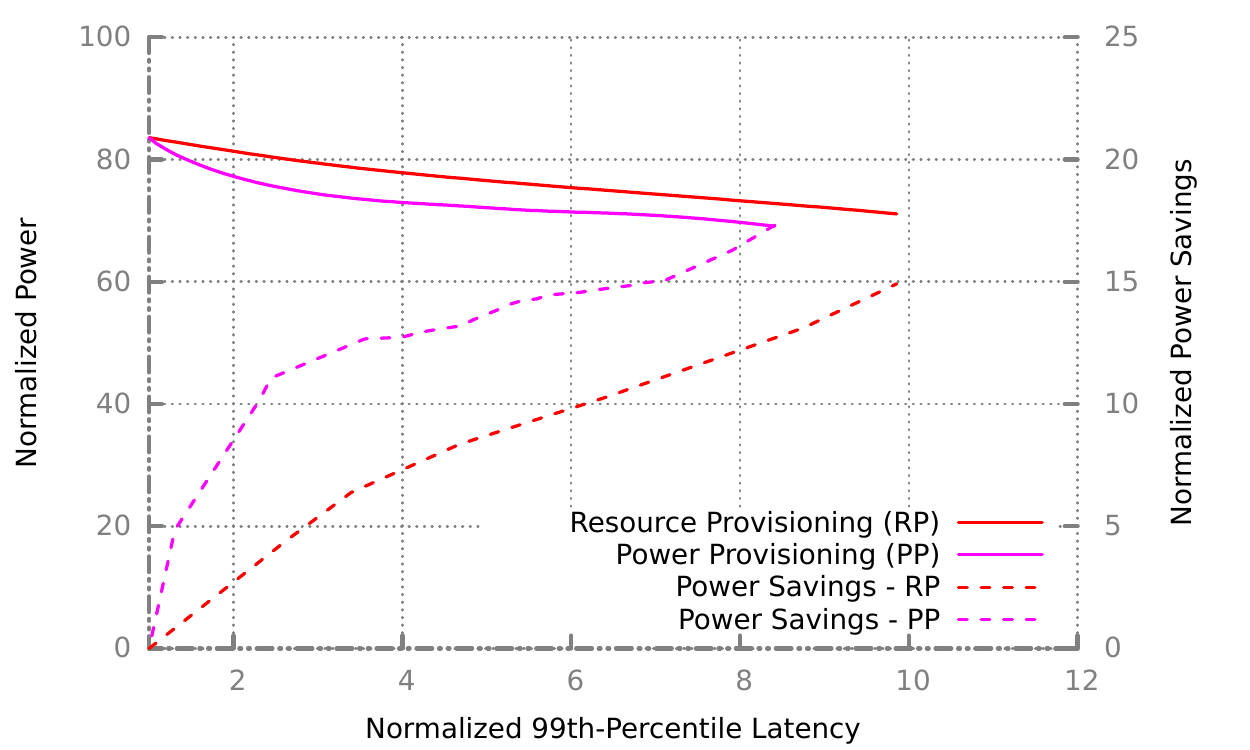}
\includegraphics[width=.329\linewidth,height=4cm]{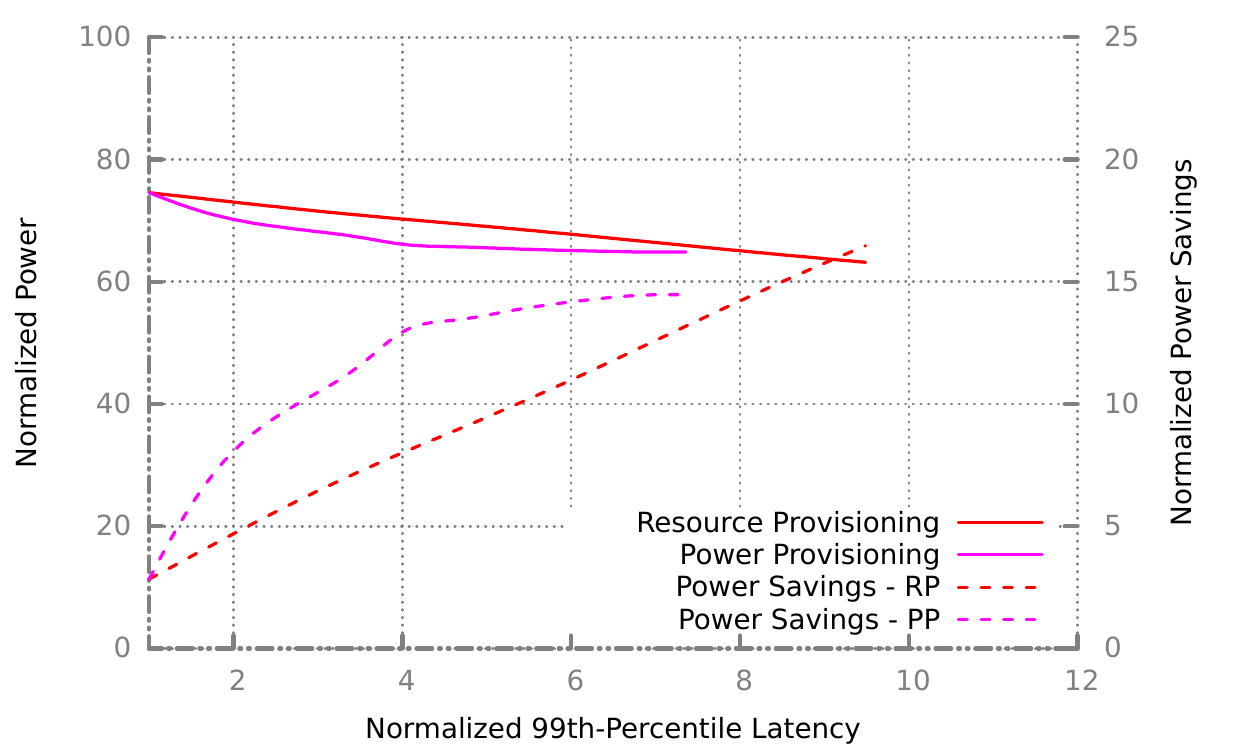}
\includegraphics[width=.329\linewidth,height=4cm]{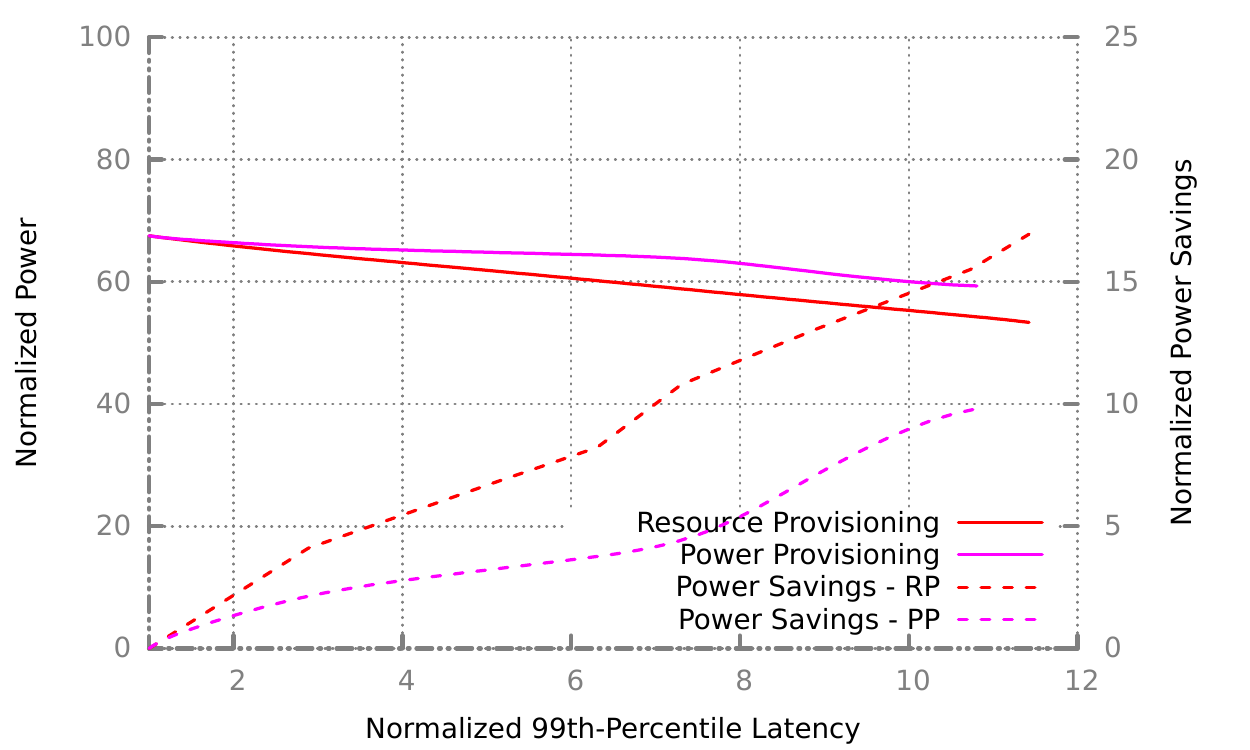}
\caption{Comparison of Power and Resource Provisioning Using Power vs. Latency 
Trade-Off. Left: 30\%, Center: 40\%, Right: 50\% Load-Levels of Data Serving Workload)}
\label{fig:ppvrp}
\end{figure*}

\subsection{Experimental Setup and Workload Configuration}

For the experiments in this section, we use a different cluster. 
32 nodes from the Shadowfax cluster housed in Virginia Tech is used as 
the evaluation testbed. Each node consists of two Intel Xeon E5-2670 processors, 
64~GB of memory and a 1~TB hard disk. A separate server
runs the workload generator or the client.

We evaluate only the data serving workload for the experiments in this section. 
Cassandra (version 2.0.7) is used as the NoSQL data store. To generate the workload for our experiments, 
YCSB (version 0.1.4) is used. We load 300 million records into the data store with
a replication factor of nine. The total data stored was 
approximately 2.85~TB in size (90~GB per node). We evaluate this setup 
using a predefined read-modify-write workload from YCSB. 
The data access pattern follows a Zipfian distribution. The arrival rate of requests follows a 
exponential distribution. 

\subsection{Methodology}

To understand the effects of resource provisioning, we change the number of 
servers involved in our experiments at each load-level from 25 to 32 nodes and report the 
effects of each configuration on the power and performance (in terms of latency). 
We first start the client with all 
the 32 nodes on the Cassandra cluster operational. After the client starts receiving responses, 
we remove the required number of nodes from Cassandra cluster. This methodology ensures 
that we use a realistic setup as we take the effects of removing a node from the Cassandra 
cluster on power and performance into account. To analyze the effects of 
power provisioning, we perform the same experiment used earlier in this paper. 
We run through different possible RAPL processor power limits and present the 
impact of each configuration on the power and performance of the data serving workload.

\subsection{Discussion}

In this section, we run the data serving workload (Cassandra) at 30, 40 and 50 percent 
load-levels. Figure~\ref{fig:ppvrp} shows the power vs latency trade-off for the 
load-levels using power and resource provisioning. The left Y-axis shows the 
power normalized to the power consumed at 100\% load-level using all the 32 servers. 
We also show the power saving using the right Y-axis. 

We are not able to achieve energy proportionality for the three load-levels presented in 
the figure. However, there is a good power saving potential depending upon how much 
performance (latency) we are willing to sacrifice. These plots can be used to make 
other interesting observations. For example, the plots show that the best 
power saving technique for a given latency depends on the load-level. The best power 
saving technique for 50\% load-level is power provisioning. But for 30\% load-level it 
is resource provisioning. In fact, the best 
power saving technique gradually shifts from power provisioning to resource provisioning, 
moving from 50\% to 30\% load-level. The reason for such a trend is that at low load-levels 
the idle power becomes a large portion of the cluster power consumption 
and active low-power modes are not capable of improving the power savings further. At low load-levels, 
the only way to improve energy-proportionality is to use resource provisioning to offset the 
idle power with an associated latency cost.

\section{Related Work}
\label{sec:related}

We give an overview of the related work in this section. 

\subsection{Energy Proportionality and Energy Efficiency} 
In our previous work~\cite{ccgrid_eprop,icpe_eprop}, we studied
the effects of RAPL power limiting on the performance, energy
proportionality and energy efficiency of enterprise applications at the 
server-level. We also designed a runtime system to decrease the energy 
proportionality gap. To design this runtime system, we used a 
load-detection model and
optimization framework that uses statistical models for capturing the
performance of an application under power limit. This paper improves 
upon our previous work by investigating the energy proportionality of 
clusters executing scale-out workloads. 

Lo et al.~\cite{pegasus} analyze the energy proportionality of web search and 
in-memory key-value data store on a large-scale cluster. They propose a runtime 
system which uses RAPL to improve the energy proportionality of these workloads.
Our work complements their paper by providing more analysis w.r.t. CPU 
utilization, turbo and active low-power modes, trade-offs in power and latency and 
the comparison of power and resource provisioning. 

Meisner et al.~\cite{oldi} characterize online data-intensive services 
(OLDI) to identify opportunities for power management, design a framework that predicts
the performance of OLDI workloads and investigate the power and
performance trade-offs using their power models and simulation framework. Our work provides 
empirical evidence for the existence of power-performance trade-offs 
in scale-out workloads on a \emph{real} system. 

Wong et al.~\cite{hpca_knightshift,knightshift} provide an infrastructure for
improving the energy proportionality using server-level
heterogeneity. They combine a high-power compute node with a low-power
processor essentially creating two different power-performance
operation regions. They save power by redirecting requests to the
low-power processor at low request rates thereby improving energy
proportionality. In addition, they compare cluster-level packing
techniques (resource provisioning) and server-level low power modes to
identify if one of these technique is better with current generation
of processors using simulation and power modeling. 
The results in this paper are applicable to commodity servers, one that does 
not require any additional hardware setup. We also provide empirical 
results for the trade-offs involved in power and resource provisioning 
on a \emph{real} system.

Fan et al.~\cite{provisioning} study the improvements
to peak power consumption of a group of servers due to the
improvements in non-peak power efficiency using their power
model. They provide analytical evidence that shows energy-proportional
systems will enable improved power capping at the data-center
level. This paper complements their work by providing more insights into 
the characteristics of scale-out workloads and the effects of power 
and resource provisioning on their energy proportionality. 

Dimitris et al.~\cite{database_node} 
provide a comprehensive study
of power consumption of relational databases on a
single node. They analyze the energy efficiency of database servers
using different hardware and software knobs such as CPU frequency,
scheduling policy and inter-query parallelism. They conclude that the
most energy-efficient operating point is also the highest performing
configuration. Willis et al.~\cite{database_cluster} study the trade-offs
between performance scalability and energy efficiency for relational
databases.  They identify hardware and software bottlenecks that
affect performance scalability and energy efficiency. In addition,
they provide guidelines for energy-efficient cluster design in the
context of parallel database software. Our research complements
theirs by addressing the energy proportionality of
non-relational (a.k.a. NoSQL) databases.

\subsection{Subsystem-Level Power Management} 

Deng et al.~\cite{coscale, memscale} 
propose the CoScale framework, which dynamically adapts the frequency of the CPU and memory 
while respecting a certain application performance degradation target.  They
also take per-core frequency settings into account. Li et
al.~\cite{crosscomponent} study the CPU microarchitectural adaptation
and memory low-power states to reduce energy consumption of
applications bounding the performance loss by using a slack allocation
algorithm.  Sarood et al.~\cite{hpc_rapl} present an interpolation
scheme to optimally allocate power for CPU and memory subsystems in an
over-provisioned high-performance computing cluster for scientific
workloads.  This paper deals with improving energy efficiency of the
compute nodes across different levels of utilization (and not just at
the peak utilization levels) as data centers running even well-tuned
applications spend a significant fraction of their time below peak
utilization levels~\cite{eprop,bubbleup,provisioning}.

\section{Conclusion}
\label{sec:conclusion}

We evaluate the potential of power and resource provisioning to improve the 
energy proportionality for scale-out workloads. Using data serving, 
web searching and data caching as our representative workloads, we show 
that processor is still the dominant power consuming component. We 
illustrate that there is ample opportunity to save power by characterizing 
the CPU utilization of scale-out workloads. We then present 
the potential of power provisioning techniques 
to improve the energy proportionality of scale-out workloads. 
The ability of active low-power modes to provide a power-performance 
trade-off space for scale-out workloads is described. We also compare and contrast power 
and resource provisioning techniques. Our study shows that effective power 
provisioning improves the energy proportionality of scale-out workloads and 
exposes the trade-off involved in power and resource provisioning.

\bibliographystyle{abbrv}	
\bibliography{eprop-scaleout}

\end{document}